\documentclass[sigconf,anonymous=false]{acmart}
\renewcommand\footnotetextcopyrightpermission[1]{} 
\AtBeginDocument{%
  \providecommand\BibTeX{{%
    \normalfont B\kern-0.5em{\scshape i\kern-0.25em b}\kern-0.8em\TeX}}}

\usepackage[capitalise]{cleveref}
\usepackage[inline]{enumitem}
\usepackage{tcolorbox}
\usepackage{pifont}
\usepackage{adjustbox}
\usepackage{xifthen}
\usepackage{mdframed}
\usepackage{todonotes}
\usepackage{amsfonts}

\usepackage{algorithm}
\usepackage[noend]{algpseudocode}

\usepackage{graphicx}
\usepackage{listings}
\usepackage{multirow}
\usepackage{booktabs}
\usepackage[skip=2pt]{caption}
\usepackage{subcaption} 
\usepackage{tikz}
\usepackage{latexsym}
\newcommand\fnote[1]{\captionsetup{font=footnotesize	}\caption*{\textmd{#1}}}

\makeatletter
\newcommand{\setword}[3]{%
  \phantomsection
  #1\def\@currentlabel{\unexpanded{#2}}\label{#3}%
}
\makeatother

\definecolor{codegreen}{rgb}{0,0.6,0}
\definecolor{codegray}{rgb}{0.5,0.5,0.5}
\definecolor{codepurple}{rgb}{0.58,0,0.82}
\definecolor{backcolour}{rgb}{0.95,0.95,0.92}

\lstdefinestyle{mystyle}{
    backgroundcolor=\color{backcolour},   
    commentstyle=\color{codegreen},
    keywordstyle=\color{magenta},
    numberstyle=\tiny\color{codegray},
    stringstyle=\color{codepurple},
    basicstyle=\ttfamily\footnotesize,
    breakatwhitespace=false,         
    breaklines=true,                 
    captionpos=b,                    
    keepspaces=true,                 
    numbers=left,                    
    numbersep=5pt,                  
    showspaces=false,                
    showstringspaces=false,
    showtabs=false,                  
    tabsize=2,
    language=C
}

\tikzset{>=latex}

\definecolor{mygrey}{RGB}{230, 230, 230}
\definecolor{bordercolor}{RGB}{190, 190, 190}

\newmdenv[topline=false,rightline=false,bottomline=false, backgroundcolor=mygrey, linecolor=bordercolor, linewidth=2]{leftbot}

\newcommand{\myparagraph}[1]{\medskip \noindent{\bf #1.}}

\def\Syntia/{Syntia}
\def\Xyntia/{Xyntia}
\def\QSynth/{QSynth}
\def\EXPR/{\textsc{Expr}}
\def\FULL/{\textsc{Full}}
\def\MBA/{\textsc{Mba}}
\def\stokesynth/{STOKE-synth}
\def\stokeopti/{STOKE-opti}

\def\hyph{-\penalty0\hskip0pt\relax}

\def\XDef/{\Xyntia/\(_{\textsc{Opt}}\)}

\newcommand{\datasetUrl}{\textbf{Will be made available}}

\def\doubleplus{+\kern-1.3ex+\kern0.8ex}

\crefname{lstlisting}{Listing}{Listings}
\Crefname{lstlisting}{Listing}{Listings}

\begin{document}

\author{Grégoire Menguy}
\email{gregoire.menguy@cea.fr}
\affiliation{%
  \institution{CEA LIST}
  \country{France}
}
\author{Sébastien Bardin}
\email{sebastien.bardin@cea.fr}
\affiliation{%
  \institution{CEA LIST}
  \country{France}
}
\author{Richard Bonichon}
\email{richard.bonichon@nomadic-labs.com}
\affiliation{%
  \institution{Nomadic Labs}
  \country{France}
}
\author{Cauim de Souza Lima}
\email{cauimsouza@gmail.com }
\affiliation{%
  \institution{CEA LIST}
  \country{France}
}

\title{AI-based Blackbox Code Deobfuscation\\ Understand, Improve and Mitigate}

\renewcommand{\shortauthors}{Menguy, et al.}

\begin{abstract}
    Code obfuscation aims at protecting Intellectual Property and other secrets 
  embedded into software from being retrieved. 
  Recent works leverage advances
  in artificial intelligence with the hope of  getting blackbox  deobfuscators completely  % 
  immune to standard (whitebox) protection mechanisms. While promising, this new field of {\it AI-based blackbox deobfuscation} is still in its infancy.   
  In this article we deepen the state of AI-based blackbox deobfuscation in three key directions: {\it understand} the current state-of-the-art, {\it improve} over it and design dedicated {\it protection mechanisms}.  
  In particular, we define a novel generic 
framework for AI-based blackbox deobfuscation encompassing  prior work and highlighting key components;  
we are the first to point out that the search space underlying code deobfuscation is too unstable for simulation-based methods (e.g., Monte Carlo Tres Search used in prior work)  and advocate the use of robust methods such as S-metaheuritics;  %  
we propose the new  optimized AI-based blackbox deobfuscator \Xyntia/ which significantly outperforms prior work in terms of success rate (especially with small time budget)  while being completely immune to the most recent anti-analysis code obfuscation methods;  
and finally we propose two novel protections against AI-based blackbox deobfuscation, allowing to counter  \Xyntia/'s powerful attacks. 

\end{abstract}


\settopmatter{printfolios=true}

\maketitle
\pagestyle{plain}
\fancyfoot{}
\thispagestyle{empty}

\section{Introduction}
\myparagraph{Context} Software contain valuable assets, such as secret
algorithms, business logic or cryptographic keys, that attackers may try to retrieve.  The so-called Man-At-The-End-Attacks scenario (MATE)
considers the case where  software users themselves are adversarial  and try to extract such information from the code.
{\it Code obfuscation} \cite{Collberg:2009:SSO:1594894,collberg1997taxonomy} aims at protecting codes against such attacks, by transforming a sensitive program $P$ into a functionally equivalent program $P'$  that is more
``difficult'' (more expensive, for example, in money or time)  to understand or modify. 
On the flip side, {\it code deobfuscation} aims to extract information from obfuscated codes.

{\it Whitebox} deobfuscation techniques, based on advanced symbolic program analysis,  have proven extremely powerful against standard obfuscation
schemes~\cite{Schrittwieser:2016,DBLP:conf/sp/BardinDM17,YadegariJWD15,DBLP:series/ais/BrumleyHLNSY08,SalwanBarPot18,DBLP:conf/wcre/Kinder12,
  BanescuCGNP16} 
  -- especially in local attack scenarios where the attacker analyses pre-identified parts of the code (e.g., trigger conditions).
  But  they  are  inherently sensitive to the {\it syntactic complexity} of the code under analysis, leading to recent and 
 effective countermeasures \cite{ollivier2019kill, ollivier2019obfuscation,Zhou:2007:IHS:1784964.1784971,Collberg:2009:SSO:1594894}.

\myparagraph{AI-based blackbox deobfuscation} Despite being rarely sound
or complete, \textit{artificial intelligence} (AI) techniques  are flexible and often provide
good enough solutions to hard problems in reasonable time. 
They  have been therefore recently applied to binary-level code deobfuscation.  
The pioneering work by Blazytko et al.~\cite{Blazytko:2017:SSS:3241189.3241240}
shows how \textit{Monte Carlo Tree
Search} (MCTS) \cite{browne2012survey} can be leveraged to solve local deobfuscation tasks by {\it learning} the semantics of pieces of protected codes in a {\it blackbox manner},  in principle {\it immune to the syntactic complexity} of these codes.  
Their method and prototype, \Syntia/, 
have been successfully used to reverse state-of-the-art protectors like VMProtect \cite{VMProtect}, Themida \cite{Themida} and Tigress \cite{Tigress},  drawing attention from the software security
community \cite{syntiaBlackHat}.

\myparagraph{Problem} While promising, AI-based blackbox (code) deobfuscation techniques are still not well understood. Several key
questions of practical relevance (e.g., deobfuscation correctness and quality,
sensitivity to time budget) are not addressed in Blazytko et al.'s original paper,  making it hard to
exactly assess the strengths and weaknesses of the approach. Moreover,  as \Syntia/ comes with many hard-coded design and implementation choices, it is legitimate to ask whether other choices  lead to better performance, and to get a broader view of AI-based blackbox deobfuscation methods. Finally, it is unclear how these methods compare with recent proposals for greybox deobfuscation \cite{davidqsynth} or general program synthesis \cite{BCD+11,DBLP:journals/corr/abs-1211-0557}, and how to protect from such blackbox attacks.    

\myparagraph{Goal} We focus on advancing the current state  of AI-based blackbox deobfuscation methods in the following three key directions: (1) generalize the initial \Syntia/ proposal and refine the initial experiments by Blazytko et al.~in order to better {\it understand} 
AI-based blackbox methods, (2) {\it improve}  the current state-of-the-art (\Syntia/) through a careful formalization and exploration of the design space and evaluate the approach against greybox and program synthesis methods,
 and finally (3)  study how to {\it mitigate} such AI-based attacks. 
 Especially, we study the underlying search space, bringing new insights 
 for efficient blackbox deobfuscation, and promote the application of S-metaheuristics \cite{Talbi:2009:MDI:1718024} instead of MCTS. 

\newpage
\myparagraph{Contributions} Our main contributions are the following:
\begin{itemize}
    \item We refine experiments by Blazytko et al. in a {\it systematic way}, highlighting both {\it new strengths and new weaknesses} of the initial \Syntia/ proposal for AI-based blackbox deobfuscation  (\cref{sec:lesson_syntia}). 
    Especially, \Syntia/ (based on Monte Carlo Search Tree, MCTS) is far less efficient than expected for small time budget (its typical usage scenario) and lacks robustness;  

    \item We propose a missing
    {\it formalization of blackbox deobfuscation} (\cref{sec:lesson_syntia}) and  dig into \Syntia/ internals to rationalize our observations (\cref{sec:syntia_explaination}). It appears that {\it the search space underlying blackbox code deobfuscation is too unstable} to rely on MCTS -- especially assigning a score to a {\it partial node} through {\it simulation} leads here to poor estimations. As a result, \Syntia/ is here {\it almost enumerative}; 
   
    \item  We propose to see  (\cref{sec:xyntia})  blackbox deobfuscation as an {\it optimization problem} rather than a {\it single player game}, allowing to reuse {\it S-metaheuristics}
    \cite{Talbi:2009:MDI:1718024}, known to be more robust than MCTS on unstable search space (especially, they do not need to score partial states). We propose \Xyntia/ (\cref{sec:xyntia}), an {\it AI-based blackbox  deobfuscator} using {\it Iterated Local Search} (ILS) \cite{lourencco2019iterated}, known among S-metaheuristics for its robustness. 
        Thorough experiments  show that  \Xyntia/ keeps the benefits of \Syntia/ while correcting most of its flaws. Especially,    
         \Xyntia/ {\it significantly outperforms} \Syntia/, 
         synthesizing  twice more expressions  with a budget of 1s/expr~than \Syntia/ with 600s/expr.  
         Other meta-heuristics also   
         clearly beat MCTS, even if they are less effective here than ILS;    

      \item We evaluate \Xyntia/ against
        other {\it state-of-the-art attackers} (\cref{sec:comparative_study}), namely the 
        \QSynth/ greybox deobfuscator~\cite{davidqsynth}, program synthesizers (CVC4 \cite{BCD+11} and STOKE~\cite{DBLP:journals/corr/abs-1211-0557}) and
        pattern-matching based simplifiers. 
        \Xyntia/ outperforms all of them -- 
        it finds $2\times$ more
        expressions and is \(30\times\) faster than \QSynth/ on heavy protections;  
        
        \item We evaluate \Xyntia/ against {\it state-of-the-art defenses}   (\cref{sec:deobfuscation_xyntia}), especially recent anti-analysis proposals \cite{Zhou:2007:IHS:1784964.1784971,collberg1998manufacturing,ollivier2019kill,10.1145/2810103.2813663,collbergprobabilistic}. 
        As expected, \Xyntia/ is immune to such defenses. In particular, it successfully bypasses side-channels \cite{collbergprobabilistic},  path explosion \cite{ollivier2019kill} and MBA  \cite{Zhou:2007:IHS:1784964.1784971}.   
        We also use it to synthesizes VM-handlers from  state-of-the-art virtualizers \cite{10.1145/2810103.2813663,Tigress,VMProtect};

      \item Finally, we propose the {\it two first protections against AI-based blackbox
        deobfuscation} (\cref{sec:protections}). We observe that all  phases of blackbox techniques can
        be thwarted (hypothesis, sampling and learning) and propose two
        practical methods exploiting these limitations, and  discuss them in the context of
        virtualization-based obfuscation: 
        \begin{enumerate*}
        \item {\it semantically complex handlers};
        \item {\it merged handlers with branch-less conditions}.
        \end{enumerate*}
        Experiments  show that both protections are highly effective against blackbox attacks. 
\end{itemize}

 \noindent We hope that our results will help better understand AI-based code
 deobfuscation, and lead to further progress in this promising field.\vspace{-1mm}
 
\myparagraph{Availability} {\it Benchmarks and code are available online\footnote{\datasetUrl{}}.
Also, we put a fair amount of experimental data in appendices for convenience. While the core paper can be read without, this material will still be made available online in a technical report.  }

\section{Background} \label{sec:background}

\subsection{Obfuscation}
\label{sec:obfuscation}
Program obfuscation \cite{Collberg:2009:SSO:1594894,collberg1997taxonomy} is a family of methods designed 
to make reverse engineering (understanding programs' internals) hard. It is
employed  by manufacturers to protect intellectual property and by malware authors to hinder analysis. It  transforms a program $P$ in a functionally  equivalent, more complex program $P'$ with an acceptable performance penalty. Obfuscation does not ensure that a program cannot be understood -- this is impossible in the MATE context \cite{barak2012possibility} -- but aims to delay 
the analysis as much as possible in order to make it unprofitable. Thus, it is especially important to protect from {\it automated deobfuscation analyses} (anti-analysis obfuscation). We
present here two important obfuscation methods.

\textbf{Mixed Boolean-Arithmetic (MBA) encoding} \cite{Zhou:2007:IHS:1784964.1784971} transforms an arithmetic and/or Boolean expression into an equivalent one, combining arithmetic and Boolean operations.
It can be applied iteratively to increase the syntactic complexity of
the expression. Eyrolles et al.~\cite{DBLP:conf/ccs/EyrollesGV16} shows that SMT solvers
struggle to answer equivalence requests on MBA expressions, preventing
the automated simplification of  protected expressions by symbolic methods. 

\begin{figure}[htbp]
\centering
\begin{center}
        \begin{adjustbox}{max width=0.8\columnwidth}
            \begin{tikzpicture}%[level/.style={sibling distance=30mm/#1, level distance=1.2cm}]

\node[minimum width=2cm,minimum height=1cm, fill=white, drop shadow] at (-3, 0) [draw, thick] (fetch) {\large Fetch};

\node[minimum width=2cm,minimum height=1cm] at (-3, -2) [align=center] (bytecodes) {\large Bytecodes};

\node[minimum width=2cm,minimum height=1cm, fill=white, drop shadow] at (0, 0) [draw, thick] (decode) {\large Decode};

\node at (0, -2) (a) {};
\node[minimum width=0,minimum height=0] at (6, -2) (b) {};

\node[minimum width=2cm,minimum height=1cm, fill=white, drop shadow] at (3, 0) [draw, thick] (exec) {\large Execute};

\node[minimum width=2cm,minimum height=3cm, fill=white, drop shadow] at (6, 0) [draw, align=center, thick] (table) {};
\node[minimum width=2cm,minimum height=3cm] at (6, -0.25) [align=center] (handlers) {$h_1(x, y)$ \\ $h_2(x, y)$ \\ $h_3(x, y)$ \\ ... \\ $h_n(x, y)$};
\node[minimum width=2cm,minimum height=0.5cm, fill=gray, text=white] at (6, 1.25) [draw, thick] (tabletitle) {\large Handlers};

\draw [->] (fetch) edge (decode);
\draw [<->] (fetch) edge (bytecodes);
\draw [->] (decode) edge (exec); 

\draw[<->] (decode) -- ++(0, -2) -| (table);
\draw[->] (exec) -- ++(0, 1.5) -| (fetch);

\end{tikzpicture}
        \end{adjustbox}
    \end{center}
    \vspace*{-2mm}
\caption{Virtualization based obfuscation}
\label{fig:vmhandlers}
\end{figure}
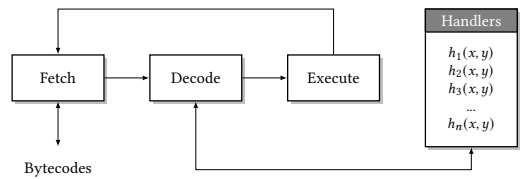

\textbf{Virtualization} \cite{10.1145/2810103.2813663}
translates an initial code
$P$ into a bytecode $B$ together with a   custom virtual machine. 
Execution of the obfuscated code can be divided in 3
steps (\cref{fig:vmhandlers}): 
\begin{enumerate*}
\item \label{vm:fetch} {\it fetch} the next bytecode instruction to execute, 
\item \label{vm:decode} {\it decod} the bytecode and finds the corresponding
  {\it handler},
\item \label{vm:execute} and finally {\it execute} the handler. 
\end{enumerate*}
Virtualization 
hides the real control-flow-graph (CFG) of $P$, and reversing the handlers is  key for reversing the VM. 
Virtualization   is notably used in malware \cite{falliere2009inside,toradevirt}.

\subsection{Deobfuscation}
\label{sec:deobfuscation}

Deobfuscation aims at reverting an
obfuscated program back to a form close enough to the original one, or at least to a more
understandable version.
Along the previous years, \textit{symbolic deobfuscation methods} based on advanced program analysis techniques have proven to be very efficient at breaking standard
protections
\cite{Schrittwieser:2016,DBLP:conf/sp/BardinDM17,YadegariJWD15,DBLP:series/ais/BrumleyHLNSY08,SalwanBarPot18,DBLP:conf/wcre/Kinder12,
  BanescuCGNP16}.  However, very effective countermeasures start to emerge, based on deep limitations of the underlying code-level reasoning mechanisms and potentially strongly  limiting their usage  \cite{ollivier2019kill,ollivier2019obfuscation,collbergprobabilistic,10.1145/2810103.2813663,BanescuCGNP16}. Especially, all such methods are ultimately {\it sensitive to the syntactic complexity} of the code under analysis.

\subsection{\Syntia/ an AI-based blackbox deobfuscator} \label{sec:back_syntia}

\textit{Artificial intelligence based blackbox deobfuscation} has been recently proposed by 
Blazytko et al. \cite{Blazytko:2017:SSS:3241189.3241240}, implemented in  the \Syntia/ tool, to learn the semantic of well-delimited code fragments, e.g.~  
MBA expressions or VM handlers. 
The code under analysis is seen as a {\it blackbox} that can only be queried (i.e., executed under chosen inputs to observe results).
\Syntia/ samples input-output (I/O) relations, then use a learning engine to find an expression mapping sampled inputs to their 
observed  outputs. Because it relies on a limited number of samples, 
results are not guaranteed to be correct. However, being fully
blackbox, it is in principle {\it insensitive to syntactic complexity}. 

\myparagraph{Scope} \Syntia/  tries to  infer a simple  semantics of {\it heavily obfuscated local code fragments} -- e.g., trigger based conditions or VM handlers. 
Understanding these fragments is critical to fulfill analysis.   

\myparagraph{Workflow} \Syntia/'s workflow is representative of AI-based blackbox deobfuscators. First, it needs  
\begin{enumerate*}
    \item a {\it reverse window} i.e., a subset of code to work on;
    \item the location of its {\it inputs} and {\it outputs}.
\end{enumerate*}
 Consider the code in \cref{lst:syntia_toy_ex} evaluating a condition at line 4. To understand this condition, a reverser focuses on the code between lines 1 and 3. This code segment is our reverse window. The reverser then needs to locate relevant inputs and outputs. The condition at line 4 is performed on $t3$. This is our output. The set of inputs contains any variables (register or memory location at assembly level) influencing the outputs. Here, inputs are $x$ and $y$. 
 Armed with these information, \Syntia/ samples inputs randomly and observes resulting outputs. In our example, 
 it might consider samples $(x \mapsto 1, y \mapsto 2)$, $(x \mapsto 0, y \mapsto 1)$ and $(x \mapsto 3, y \mapsto 4)$ which respectively evaluate $t3$ to $3$, $1$ and $7$. \Syntia/ then synthesizes an expression matching these observed behaviors, using  Monte Carlo Tree Search (MCTS) over the space of all possible (partial) expressions.   
 Here, it rightly infers that $t3 \gets x + y$ and  the reverser concludes that the condition is $x + y = 5$, 
 where a symbolic method
 will typically simply retrieve that $  ((x \lor 2y) \times 2 - (x \oplus 2y) - y) = 5$.

 \begin{lstlisting}[basicstyle=\footnotesize,caption={Obfuscated condition},label={lst:syntia_toy_ex}]
int t1 = 2 * y;
int t2 = x | t1;
int t3 = t2 * 2 - (x ^ t1) - y;
if (t3 == 5) ...
\end{lstlisting}

\section{Motivation} \label{sec:motiv_ex}

\subsection{Attacker model}
\label{sec:attacker-model}

In the MATE scenario, the attacker is the software user himself. 
He has only access to the obfuscated version of the code under analysis and can read or run it at will.
We consider that the attacker is highly skilled in reverse engineering but has limited resources in terms of time or money. 
We see reverse engineering as a {\it human-in-the-loop} process  where the attacker combines manual analysis with  automated state-of-the-art deobfuscation methods  (slicing, symbolic execution, etc.) on critical, heavily obfuscated code fragments  like VM handlers or trigger-based conditions. 
Thus, an effective defense strategy is to thwart automated deobfuscation methods. %

\subsection{Syntactic and semantic complexity}
\label{sec:use-black-box}

We now intuitively motivate the use of blackbox deobfuscation.
Consider that we reverse a software protected through virtualization. We need to extract the semantics of all handlers, which usually perform basic operations like $h(x, y) = x+y$. Understanding $h$ is trivial, 
but it can be protected  to hinder analysis.
\cref{eq:example_mba} shows how MBA encoding hides $h$'s semantics.
\begin{equation} \label{eq:example_mba}
    h(x, y) = x + y \hspace{2mm} \overset{mba}{\longrightarrow} \hspace{2mm} (x \lor 2y) \times 2 - (x \oplus 2y) - y 
\end{equation}

Such encoding {\it syntactically} transforms the expression to make it
incomprehensible while preserving its {\it semantics}. To highlight the difference between
syntax and semantics, we distinguish:
\begin{enumerate}
    \item \setword{\textbf{The syntactic complexity}}{Def.}{syntac_compl} of  expression $e$ is the size of~$e$, i.e.~the number of operators used in it;
 
    \item \textbf{The semantic complexity} of  expression $e$  is the smallest size of expressions $e'$ (in a given language) equivalent to $e$. 
\end{enumerate}
For example, in the MBA language, $x + y$ is syntactically simpler than $(x \lor 2y) \times 2 - (x
\oplus 2y) - y$, yet  they have the same semantic complexity as they are equivalent. Conversely, \(x+y\) is more semantically complex than $(x+y)
\land 0$, which equals $0$. 
We do not claim to give a definitive definition of semantic and syntactic
complexity -- as smaller is not always simpler -- but introduce the
idea that two kinds of complexity exist and are independent.

The encoding in \cref{eq:example_mba} is simple, but it can be repeatedly applied to create a more syntactically complex expression, leading the reverser to either give up or try to simplify it automatically. Whitebox methods based on \textit{symbolic execution} (SE) \cite{SalwanBarPot18,YadegariJWD15} and {\it formula simplifications} (in the vein of compiler optimizations) can extract the semantic of an expression, yet 
they  are sensitive to syntactic complexity and will not return simple versions of highly obfuscated expressions. 
Conversely, \textit{blackbox deobfuscation} treats the code as a blackbox, considering only sampled I/O behaviors. 
{\it Thus increasing syntactic complexity, as usual state-of-the-art protections do, has simply no impact on blackbox methods}. 

\subsection{Blackbox deobfuscation in practice} \label{sec:bb_practice}

We now present how blackbox methods integrate in a global deobfuscation process and highlight crucial properties they must hold. 

\myparagraph{Global workflow} Reverse engineering can be fully automated, 
or handmade by a reverser, leveraging tools to automate specific tasks.
While the deobfuscation process operates on the whole obfuscated binary, blackbox modules can be used to analyze parts of the code like conditions or VM handlers.  Upon meeting a complex code fragment, the blackbox deobfuscator is called to retrieve a simple semantic expression. After synthesis succeeds, the inferred expression is used to help continue the analysis. 

\myparagraph{Requirements} In virtualization based obfuscation, the blackbox module is typically  queried on all VM  handlers~\cite{Blazytko:2017:SSS:3241189.3241240}. As the number of handlers can be arbitrarily high, blackbox methods need to be {\it fast}. 
In addition, inferred expressions  should ideally  be as {\it simple} as the original non-obfuscated expression and {\it semantically equivalent} to the obfuscated expression (i.e. correct). Finally, {\it robustness} (i.e. the capacity to synthesize complex expressions) is needed to be usable in various situations. 
 Thus,  \textbf{speed}, \textbf{simplicity}, \textbf{correctness} and \textbf{robustness},  are required  for efficient blackbox deobfuscation. 

\section{Understand AI-based deobfuscation} \label{sec:lesson_syntia}

We propose a general view of AI-based code deobfuscation fitting state-of-the art solutions~\cite{Blazytko:2017:SSS:3241189.3241240,davidqsynth}. We also extend the evaluation of  \Syntia/ by Blazytko et al.~\cite{Blazytko:2017:SSS:3241189.3241240}, highlighting both some previously unreported weaknesses and strengths. From that we derive general lessons on 
the (in)adequacy of MCTS for code deobfuscation, that will guide  our new approach  (\cref{sec:xyntia}). 

\subsection{Problem at hand}

AI-based deobfuscation takes an obfuscated expression and tries to infer an equivalent one with lower syntactic complexity. Such problem can be stated as following: 

\myparagraph{Deobfuscation} Let $e$, $obf$  be 2 equivalent expressions such that $obf$ is an obfuscated version of $e$ -- note that  $obf$ is possibly much larger than $e$. Deobfuscation aims to infer an expression $e'$ equivalent 
to $obf$ (and $e$), but   with size similar to $e$.  
Such problem can be approached in three ways depending on the amount of information given to the analyzer: 

\begin{description}[wide]
    \item[Blackbox] We can only run $obf$. The search is thus driven by sampled I/O behaviors. \Syntia/~\cite{Blazytko:2017:SSS:3241189.3241240} is a blackbox approach;  
    
    \item[Greybox] Here $obf$ is executable and readable but the semantics of its operators is mostly unknown. The search is driven by previously sampled I/O behaviors
        which can be applied to subparts of $obf$. \QSynth/ \cite{davidqsynth} is a greybox solution;  
    \item[Whitebox] The analyzer has full access to $obf$ (run, read) and the semantics of its operators is precisely known. Thus, the search can profit from advanced pattern matching and symbolic  strategies. Standard static analysis falls in this category. 
\end{description}

\myparagraph{Blackbox methods} AI-based blackbox deobfuscators follow the framework given in \cref{alg:bb_frame}. In order to deobfuscate code, one must detail a {\it sampling strategy} (i.e.,
how inputs are generated), a {\it learning strategy} (i.e., how to learn an expression mapping sampled inputs to observed outputs) and a {\it simplification postprocess}. 
For example, \textbf{\Syntia/} samples inputs \textit{\textbf{randomly}}, uses
\textit{\textbf{Monte Carlo Tree Search}} (MCTS)~\cite{browne2012survey} as learning strategy and leverages the \textit{\textbf{Z3 SMT solver}} \cite{de2008z3} for  simplification.
The choice of the sampling and learning strategies is critical. For example, too few samples could lead to incorrect results while too many could impact the search efficiency, and  an inappropriate learning algorithm could impact robustness or speed.  

Let us now turn to discussing \Syntia/'s learning strategy. We show that using MCTS leads to disappointing performances and give insight to understand why.  
 
\begin{algorithm}
\caption{AI-based blackbox deobfuscation framework}\label{alg:bb_frame}
    \begin{flushleft}
    \textbf{Inputs:}\\
    \hspace*{\algorithmicindent}$Code$ : code to analyze \\
    \hspace*{\algorithmicindent}$Sample$ : sampling strategy \\
    \hspace*{\algorithmicindent}$Learn$ : learning strategy \\
    \hspace*{\algorithmicindent}$Simplify$ : expression simplifier \\
    \textbf{Output:} learned expression or Failure \\
    \end{flushleft}
\begin{algorithmic}[1]
\Procedure{Deobfuscate}{$Code, Sample, Learn$}
    \State $Oracle \gets Sample(Code)$
    \State $succ,\; expr \gets Learn(Oracle)$
    \If{$succ = True$}
        \textbf{return} $Simplify(expr)$
    \Else
        \textbf{ return} $Failure$
    \EndIf
\EndProcedure
\end{algorithmic}
\end{algorithm}

\subsection{Evaluation of \Syntia/} \label{sec:syntia_eval}

We extend \Syntia/'s evaluation and tackle the following questions left unaddressed by Blazytko et al.~\cite{Blazytko:2017:SSS:3241189.3241240}. 

\newlist{RQ}{enumerate}{1}
\setlist[RQ]{label={\bf RQ\arabic*}, ref={\bf RQ\arabic*}, leftmargin=* ,parsep=0cm,itemsep=0cm,topsep=0cm, align=left}
\begin{RQ}[resume]
    \item \label{qq1} \emph{Are results stable across different runs?} \\
        This is desirable  due to the stochastic nature of MCTS; 
    \item \label{qq2} \emph{Is \Syntia/ fast, robust and does it infer simple and correct results?} \\
        \Syntia/ 
        offers {\it a priori} no guarantee of correctness nor quality. Also, we consider small time budget (1s), adapted to human-in-the-loop reverse scenarios but absent from the initial evaluation;
        
    \item \label{qq3} \emph{How is synthesis impacted by the set of operators' size?} \\
        \Syntia/ learns expressions over a search space fixed by
        predefined grammars. Intuitively, the more operators in
        the grammar, the harder it will be to converge to a solution. We use 3 sets of operators to assess this impact.
 
\end{RQ}

\subsubsection{Experimental setup} We distinguish 
the {\bf success rate} (number of expressions inferred) from 
the {\bf equivalence rate} (number of expressions inferred and equivalent to
the original one). The equivalence rate relies on the Z3 SMT solver~\cite{de2008z3} 
with a timeout of 10s. 
Since Z3 timeouts are inconclusive answers, we
define a notion of \textbf{equivalence range}:  its lower bound is the
\textbf{proven equivalence rate} (number of expression proven to be equivalent) while its  upper bound is the \textbf{optimistic equivalence rate}
(expressions not proven different, i.e., optimistic~=~proven~+~\#timeout). The equivalence rate is within the equivalence range, while the  success rate is higher than the optimistic equivalence rate.  Finally, we define the
{\bf quality} of an expression as the ratio between the number of operators in recovered and target expressions. It estimates the syntactic complexity of inferred expressions compared to the original ones. 
A quality of 1 indicates that the recovered expression has the same size as the target one.

\myparagraph{Benchmarks} We consider two benchmark suites: B1 and
B2. B1\footnote{https://github.com/RUB-SysSec/syntia/tree/master/samples/mba/tigress} comes from Blazytko et al.~\cite{Blazytko:2017:SSS:3241189.3241240} and was used to evaluate \Syntia/. It comprises 500 randomly generated
expressions with up to 3 arguments, and simple semantics. It aims at representing state-of-the-art VM-based obfuscators. {\it However, we found that B1 suffers from several significant issues}: 
\begin{enumerate*}[label=(\arabic*)]
    \item it is not well distributed over the number of inputs and expression types,  making it unsuitable for fine-grained analysis; 
    \item only 216 expressions are unique modulo renaming -- the other
        284 expressions are $\alpha$-equivalent, like ~x+y and a+b.
\end{enumerate*}
These problems threaten the validity of the evaluation. 

We thus {\it propose a
  new benchmark} \textbf{B2}
consisting of 1,110 
randomly generated
expressions, better distributed according to number of
inputs and nature of operators -- see \cref{annex:xyntia_experimental_eval} for details. We use three categories of
expressions: Boolean, Arithmetic and Mixed Boolean-Arithmetic, with 2
to 6 inputs. Each expression has an Abstract Syntax Tree (AST) of maximal height
3. As a result, B2 is more challenging than B1 and enables a finer-grained evaluation.

\myparagraph{Operator sets} \cref{tab:operator_sets} introduces three  operator sets: \FULL/, \EXPR/ and \MBA/. We use these to evaluate sensitivity to the search space and
answer \ref{qq3}. \EXPR/ is as expressive as \FULL/ even if \EXPR/ \(\subset\)
\FULL/. \MBA/ can only express Mixed Boolean-Arithmetic expressions~\cite{Zhou:2007:IHS:1784964.1784971}.

\begin{table}[ht]

  \centering
  \caption{Sets of operators}
  \resizebox{\columnwidth}{!}{%
  \begin{tabular}{@{}l@{ : }l@{}}
      \textbf{\FULL/} & $\{ -_1, \neg, +, -, \times, \gg_u, \gg_s, \ll, \land, \lor, \oplus, \div_s, \div_u, \%_s, \%_u, \doubleplus{} \}$  \\
     \textbf{\EXPR/} & $\{ -_1, \neg, +, -, \times, \land, \lor, \oplus, \div_s, \div_u, \doubleplus{} \}$  \\
     \textbf{\MBA/}  & $\{ -_1, \neg, +, -, \times, \land, \lor, \oplus \}$  \\
  \end{tabular}
  }

\label{tab:operator_sets}
\end{table}

\myparagraph{Configuration} We run all our experiments on a machine with 6 Intel Xeon E-2176M  CPUs and 32 GB of RAM.
We evaluate \Syntia/ in its original
configuration~\cite{Blazytko:2017:SSS:3241189.3241240}: the SA-UCT constant is
1.5, we use 50 I/O samples and a maximum playout depth of 0. It also limits
\Syntia/ to 50,000 iterations per sample, corresponding to a timeout of 60~s per
sample on our test machine.

\subsubsection{Evaluation Results}
Let us summarize here the outcome of our experiments --- see \cref{annex:lessons_syntia} for complete results. 

\myparagraph{\ref{qq1}} Over 15 runs, \Syntia/ finds between 362
and 376 expressions of B1 i.e., 14 expressions of difference ($2.8\%$ of B1). Over B2, it finds between 349 and 383 expressions i.e., 34 expressions of
difference (3.06\% of B2). Hence, \textit{\Syntia/ is very stable across executions}.

\myparagraph{\ref{qq2}} \Syntia/ cannot efficiently infer B2 ($\approx 34\%$ success rate). Moreover, \cref{tab:syntia_timeouts} shows \Syntia/ to be highly sensitive to time budget. More precisely, with a time budget of 1s/expr., \Syntia/ only retrieves 16.3\% of B2. Still, even with a timeout of 600~s/expr., it tops at $42\%$ of B2.
In addition, \Syntia/ is unable to synthesize expressions with more than 3 inputs -- success rates for 4, 5 and 6 inputs respectively falls to 10\%, 2.2\% and 1.1\%. It also struggles over expressions using a mix of boolean and arithmetic operators, synthesizing only 21\%.
 Still, \Syntia/ performs well regarding quality and correctness. On average, its quality is around 0.60 (for a timeout of 60s/expr.) i.e., resulting expressions are simpler than the original (non obfuscated) ones, and it rarely returns non-equivalent expressions -- between 0.5\% and 0.8\% of B2. 
 We thus conclude that \textit{\Syntia/ is stable and returns correct and simple results. Yet, it is  not efficient enough (solve only few expressions on B2, heavily impacted by
time budget) and not robust (number of inputs and expression's type).} 

\begin{table}[ht]
  \centering
  \caption{\Syntia/ depending on the timeout per expression (B2)}
  \resizebox{\columnwidth}{!}{%
 \begin{tabular}{lcccc}
      & 1s & 10s & 60s & 600s \\
    \cmidrule(lr){1-1}\cmidrule(lr){2-2}\cmidrule(lr){3-3}\cmidrule(lr){4-4}\cmidrule(lr){5-5}
    Succ. Rate & 16.5\% & 25.6\% & 34.5\% & 42.3\% \\
    Equiv. Range & 16.3\% & 25.1 - 25.3\% & 33.7 - 34.0\% & 41.4 - 41.6\% \\
    Mean Qual & 0.35  & 0.49 & 0.59 & 0.67 \\
  \end{tabular}
  }
  \label{tab:syntia_timeouts}
\end{table}

\myparagraph{\ref{qq3}} Default \Syntia/ synthesizes 
expressions over the \FULL/ set of operators. To evaluate its
sensitivity to the search space we run it over \FULL/, \EXPR/ and 
\MBA/. 
Smaller sets do exhibit higher success rates (42\% on \MBA/)
but results remain disappointing. \textit{ \Syntia/ is sensitive to the size of the operator set but is inefficient even with \MBA/}.

\myparagraph{Conclusion} {\it \Syntia/ is stable, correct  and returns simple results. Yet, it is heavily impacted by the time budget and lacks robustness. It thus fails to meet the requirements given in \cref{sec:bb_practice}.} 

\subsection{Optimal \Syntia/}\label{best_syntia}

To ensure the conclusions given in \cref{sec:syntia_explaination} apply to MCTS and not only to \Syntia/, we study \Syntia/ extensively to find better set ups (\cref{annex:lessons_syntia}) for the  following parameters: simulation depth, SA-UCT value (configuring the balance between exploitative and explorative behaviors), number of I/O samples and  distance.
Optimizing \Syntia/'s parameters slightly improves its results which stay disappointing (at best, $\approx 50\%$ of success rate on \MBA/ in 60~s/expr.).

\myparagraph{Conclusion} {\it By default, \Syntia/ is well configured. Changing its parameters lead in the best scenario to marginal improvement, hence the pitfalls highlighted seem to be inherent to the MCTS approach.}

\subsection{MCTS for deobfuscation}\label{sec:syntia_explaination}

Let us explore whether  these issues  are related to  MCTS.

\myparagraph{Monte Carlo Tree Search} MCTS creates here a search tree  where each node is an {\it expression} which can be {\it terminal} (e.g. $a + 1$, where $a$ is a variable) or {\it partial} (e.g. $U + a$, where $U$ is a non-terminal symbol). 
The goal of MCTS is to expand the search tree smartly, {\it focusing on most pertinent nodes first}. 
Evaluating the pertinence of a {\it terminal node} is done by {\it sampling} (computing here a distance between the evaluation of sampled input over the node expression against their expected output values). 
For {\it partial nodes}, MCTS relies on {\it simulation}: random rules of the grammar are applied to the expression (e.g., $U+a \leadsto b+a$~) until it becomes terminal and is evaluated.
As an example, let $\{(a \mapsto 1, b \mapsto 0), (a \mapsto 0, b \mapsto 1)\}$ be the sampled inputs. The expression $b+a$ (simulated from $U+a$) evaluates them to $(1, 1)$. If the ground-truth outputs are $1$ and $-1$, the distance will equal $\delta(1, 1) + \delta(1, -1)$ where $\delta$ is a chosen distance function. We call the result the {\it pertinence measure}. The closer it is to $0$, the more pertinent the node $U+a$ is considered and the more the search will focus on it.

\myparagraph{Analysis} This {\it simulation-based pertinence estimation} is not reliable in our code deobfuscation setting. 

\begin{itemize} 

\item We present in \cref{fig:simulation_dispertion}, for different non-terminal nodes, the distance values computed through simulations. 
We observe that from a starting node, a random simulation can return drastically different results. 
It shows that {\it the search space is  very unstable} and that relying on a simulation is misleading (especially in our context where time budget is small);  

\item Moreover, our experiments show that in practice \Syntia/ is not guided by simulations and behaves {\it almost as if it were an  enumerative (BFS) search} -- MCTS where simulation is non informative. As an example,
\cref{fig:syntia_dist_timeouts} compares how the distance evolves over time for \Syntia/ and a custom, fully enumerative, MCTS synthesizer: both are very similar;  

\item Finally, on B2 with a timeout of 60 s / expr, only 34/341 successfully synthesized expressions are the children of previously  most promising nodes. It shows that \Syntia/ successfully  synthesized expressions due to its exploratory (i.e., enumerative) behavior rather than  to the selection of nodes according to their {\it pertinence}. 

\end{itemize}

\begin{figure}[!ht]
    \centering
     \resizebox{\columnwidth}{!}{%
     \input{./graphs/dispertion.tex}
     }
     \vspace{-1.5cm}
    \noindent\fnote{
    Each point represents the distance between $(a \land b) \times (b + c)$ and one simulation of a non terminal expression (horizontal axis). A non terminal expression, can generate multiple terminal ones through simulations, leading to completely different results.}
    \caption{Dispersion of the distance for different simulations}
     
    \label{fig:simulation_dispertion}
\end{figure}

\begin{figure}[!ht]
    \centering
     \resizebox{0.95\columnwidth}{!}{%
     \input{./graphs/mcts_enum_time.tex}
    }
    \caption{\Syntia/ and enumerative MCTS: distance evolution}
    \label{fig:syntia_dist_timeouts}
\end{figure}

\myparagraph{Conclusion} {\it The search space from blackbox code deobfuscation is too unstable, making MCTS's simulations unreliable. MCTS in that setting is then almost enumerative and inefficient. That is why \Syntia/ is slow and not robust,  but returns simple expressions.}

\subsection{Conclusion}

 While \Syntia/ returns simple results, it only synthesizes semantically simple expressions and is slow. These unsatisfactory results can be explained by the fact that the search space is too unstable, making the use of MCTS unsuitable. In the next section, we show that methods avoiding the  manipulating of partial expressions (and thus free from simulation)  
 are better suited to deobfuscation.

\section{Improve ai-based deobfuscation} \label{sec:xyntia}

We define a new AI-based blackbox deobfuscator, dubbed \Xyntia/, leveraging
{\it S-metaheuristics} \cite{Talbi:2009:MDI:1718024} and {\it Iterated Local Search} (ILS) \cite{lourencco2019iterated} 
and compare its design to rival deobfuscators. Unlike MCTS, S-metaheuristics  {\it only manipulate  terminal expressions} and do not create  tree searches, thus we expect them to be better suited than MCTS for code deobfuscation.  Among S-metaheuristics, ILS is particularly {\it designed for unstable search spaces}, with the ability to remember the last best solution encountered and restart the search from that point.  
We show that these methods are well-guided by the distance function and significantly outperform  MCTS in the context of blackbox code deobfuscation.   

\subsection{Deobfuscation as Optimization} \label{sec:formalization}

As presented in \cref{sec:lesson_syntia}, \Syntia/ frames  
 deobfuscation as a single player game. We instead propose to frame it as an optimization problem using ILS as learning strategy.

\myparagraph{Blackbox deobfuscation: an optimization problem} Blackbox
deobfuscation synthesizes an expression from 
inputs-outputs samples and can be modeled  as an optimization
problem. The objective function, noted $f$, 
measures the similarity between current and ground truth behaviors by computing
the sum of the distances between found and objective outputs. 
The goal is to infer an expression minimizing the objective function over the I/O samples. 
If the underlying grammar is expressive enough, a minimum exists and matches all
sampled inputs to objective outputs, zeroing $f$. The reliability of the found
solution depends on the number of I/O samples considered. Too few samples
would not restrain search enough and lead to flawed results.

\myparagraph{Solving through search heuristics}  S-metaheuristics \cite{Talbi:2009:MDI:1718024} can be advantageously used to solve such optimization problems. A wide range of heuristics exists (Hill Climbing, Random Walk, Simulated Annealing, etc.). They all
iteratively improve a
candidate solution by testing its ``neighbors'' and moving along the search space.  
Because solution improvement is evaluated by the objective function, it is said to guide the search. 

\myparagraph{Iterated Local Search} Some S-metaheuristics are
 prone to be stuck in local optimums so that the result 
 depends on the initial
 input chosen. Iterated Local Search (ILS) \cite{lourencco2019iterated} tackles the problem 
 through iteration of search and the ability to restart from previously seen best solutions. 
Note that ILS is parameterized by another search heuristics (for us: Hill Climbing).  
 Once a local  optimum is found by this side search, ILS perturbs it and uses the perturbed solution as initial state for the side search.   
 At each iteration, ILS also saves the best solution found. Unlike most other S-metaheurtics (Hill Climbing, Random Walk, Metropolis Hasting and Simulated Annealing, etc.), if the search follows a misleading path, ILS can restore the best seen solution so far to restart from an healthy state.

\subsection{\Xyntia/'s internals} \label{sec:xyntia_intern}

\Xyntia/ is built upon 3 components: the \textit{optimization} problem we aim to solve,  the \textit{oracle} which extracts the sampling information 
from the protected code under analysis and the {\it search heuristics}.

\myparagraph{Oracle} The \textit{oracle} is defined by the sampling strategy which depicts how the protected program must be sampled and how many samples are considered.
As default, we consider that our oracle samples 100 inputs over the range $[-50; 49]$.
Five
are not randomly generated but equal interesting
constant vectors ($\vec{0}, \vec{1}, \vec{-1}, \vec{min_s}, \vec{max_s}$). These choices arise from a systematic study of the different settings to find 
the best design (see \cref{annex:features}).  

\myparagraph{Optimization problem} The \textit{optimization problem} is defined as follow. The search space is the set of expressions expressible using the \EXPR/ set of operators (see \cref{tab:operator_sets}),
and considers 
a unique constant \texttt{1}. 
This grammar enables \Xyntia/ to reach optimal results while being as expressive as rivals' tools like \Syntia/ \cite{Blazytko:2017:SSS:3241189.3241240}.
Besides, we consider the objective~function: 
\[
    f_{\vec{o}^{\,*}}(\vec{o}\,) = \underset{i}{\sum}\; log_2(1 + |o_i - o_i^*|)
\]

\noindent 
It computes the Log-arithmetic distance between synthesized expressions' outputs ($\vec{o}\,$) and sampled ones ($\vec{o}^{\,*}$). The choice of the grammar and of the objective function are respectively discussed in \cref{sec:experimental_eval_xyntia,annex:features}.

\myparagraph{Search} 
\Xyntia/ leverages Iterated Local Search (ILS) to minimize our objective function and so to synthesize target expressions.
We present now how ILS is adapted to our context. 
ILS applies two steps
starting from a random
terminal (a constant or a variable):

\begin{itemize}
\item ILS reuses the {\it best expression found so far} to {\it perturb it} by 
  randomly selecting a node of the AST and replacing it by a random {\it terminal} node. The resulting AST is kept even if the distance increases and passed to the next step.
\item \emph{Iterative Random Mutations:} the side search (in our case Hill
    Climbing) iteratively mutates the input expression until it cannot improve anymore. We estimate that
    no more improvement can be done after 100 inconclusive mutations.
    A mutation (see \cref{fig:ex_mutation}) consists in replacing a randomly chosen node of the abstract syntax tree (AST) by a leaf or an AST of depth one (only one operator). 
  At each mutation, it keeps the version of the AST minimizing the distance
  function. 
  During mutations, the {\it best solution so far} is updated to be restored in the perturbation step.
  If a solution nullifies the objective function, it is directly returned.

\end{itemize}

These two operations are iteratively performed until time is out (by default {\bf 60~s}) or
an expression mapping all I/O samples is found. 
Furthermore, as \Syntia/ applies Z3's simplifer to "clean up"  recovered  expressions, we add a custom {\it post-process expression simplifier}, applying simple rewrite rules until a fixpoint is reached. \cref{annex:features}  compares \Xyntia/ with and without simplification.   
\Xyntia/ is implemented in \textit{OCaml}
\cite{OcamlManual}, within the BINSEC framework for binary-level program analysis \cite{david2016binsec}. 
It comprises $\approx$9k lines of code.

\begin{figure}[ht]
  
  \resizebox{.7\columnwidth}{!}{%
\begin{tabular}{ccc}
        \midrule
        \textit{Random selection} & & \textit{Mutated} \\
      %  & & \\
        \fcolorbox{red}{white}{$1$} $+\, (- a)$ & $\overset{mutation}{\longrightarrow}$ & $(- b) + (- a)$\\
        \midrule
  \end{tabular}
  }
  \vspace{2mm}
  \caption{Random mutation example}
  \label{fig:ex_mutation}
\end{figure}

\subsection{\Xyntia/ evaluation} \label{sec:experimental_eval_xyntia}

We now evaluate \Xyntia/ in depth and compare it to \Syntia/.
As with \Syntia/ we answer the following questions:

\begin{RQ}[resume]
    \item \label{q1} \emph{Are results stable across different runs?} 
    \item \label{q2} \emph{Is \Xyntia/ robust, fast and does it infer simple and correct results?} 
        
    \item \label{q3} \emph{How is synthesis impacted by the set of operators' size?}
\end{RQ}

\def\XyntiaDefault/{\Xyntia/$_{\textsc{Opt}}$}
\myparagraph{\setword{Configuration}{\XDef/}{config}} For all our experiments,
 we default to  locally optimal \Xyntia/ (\XDef/) presented in \cref{sec:xyntia_intern}. It learns expressions over \EXPR/, samples 100 inputs (95 randomly and 5 constant vectors) and uses the Log-arithmetic distance as objective function. 
 
 {\it Interestingly, all results reported here also hold (to a lesser extend regarding efficiency) 
 for other \Xyntia/ configurations (\cref{sec:features}), especially these versions  consistently beat \Syntia/.}

\myparagraph{\ref{q1}} Over 15 runs \Xyntia/ always finds all 500 expressions in B1 and between 1051 to 1061 in B2. The difference between the best and the worst case is only 10 expressions (0.9\% of B2). Thus, \textit{\Xyntia/ 
is very stable across executions}.

\myparagraph{\ref{q2}} Unlike \Syntia/, \Xyntia/ performs well on both
B1 and B2 with a timeout of 60~s/expr. \cref{fig:syntia_vs_syntia_timeouts} reveals
that it is still successful for a timeout of 1 s/expr. (78\% proven equivalence
rate). 
Moreover, for a timeout of 600~s/expr.
(10~min), \Syntia/ finds $2\times$ fewer expressions than \Xyntia/ with a 1~s/expr. time budget. In addition, \Xyntia/ handles well expressions using up to 5 arguments and all expression types.
Its mean quality is around 0.93, which is very good (objective is 1),  
and it
rarely returns not equivalent expressions -- only between 1.3\% and 4.9\%.
Thus,
\textit{\Xyntia/ reaches high success and equivalence rate. 
It is fast, synthesizing most expressions in  \(\leq 1 s\), 
 and it returns simple and correct results.} 

\begin{figure}[!ht]
    \centering
     \resizebox{0.8\columnwidth}{!}{%
    \input{./graphs/timeouts.tex}
    }
    \caption{Equivalence range of \Syntia/ and \Xyntia/ (\ref{config}) depending on timeout (B2)}
    \label{fig:syntia_vs_syntia_timeouts}
\end{figure}

\myparagraph{\ref{q3}}  \Xyntia/ by default  synthesizes expressions over \EXPR/ while \Syntia/ infers expressions over  \FULL/. To compare their
sensitivity to search space and show that previous results was not due to search space inconsistency, we run \Xyntia/ over \FULL/, \EXPR/ and 
\MBA/ and compare it to \Syntia/. Experiments shows that \Xyntia/ reaches high
equivalence rates for all operator sets while \Syntia/ results stay low. Still,
\Xyntia/ seems more sensitive to the size of the set of operators than \Syntia/. Its proven equivalence rate decreases from 90\% (\EXPR/)
to 71\% (\FULL/) while \Syntia/ decreases only from 38.7\% (\EXPR/) to 33.7\% (\FULL/). Conversely, as for \Syntia/, restricting to \MBA/ benefits to \Xyntia/.
Thus, \textit{like \Syntia/, \Xyntia/ is sensitive to the size of the
    operator  set. 
     Yet,
    \Xyntia/ reaches high equivalence rates even on \FULL/ while \Syntia/ remains inefficient even on \MBA/}.

\myparagraph{Conclusion} {\it \Xyntia/ is a lot faster and more robust than \Syntia/. It is also stable and returns simple expressions. Thus, \Xyntia/, unlike \Syntia/, meets the requirements given in \cref{sec:bb_practice}.}

\subsection{Optimal \Xyntia/ and other S-Metaheuristics}\label{sec:features}

Previous experiments consider the \ref{config} configuration of \Xyntia/. It comes from a systematic evaluation of the design space (\cref{annex:features}). To do so, we considered
\begin{enumerate*}
    \item different S-metaheuristics (Hill Climbing, Random Walk, Simulated Annealing, Metropolis Hasting and Iterated Local Search);
    \item different sampling strategies;
    \item different objective functions.
\end{enumerate*}
This evaluation confirms that \ref{config} is locally optimal and that ILS, being able to restore best expression seen after a number of unsuccessful mutations, outperforms other S-metaheuritics.  
Moreover, all S-metaheurstics -- except Hill Climbing -- 
outperforms \Syntia/. 

It confirms that estimating non terminal expression's pertinence through simulations, as MCTS does, is not suitable for deobfuscation (\cref{sec:syntia_explaination}). It is far more relevant to manipulate terminal expressions only as S-metaheurstics.

\myparagraph{Conclusion} {\it Principled and systematic evaluation of \Xyntia/'s design space lead to the locally optimal \ref{config} configuration. It notably shows that 
ILS outperforms other tested S-metaheuristics. Moreover,
all these S-metaheuristics -- except Hill Climbing -- outperform MCTS, confirming that manipulating only terminal expressions is beneficial.}

\subsection{On the effectiveness of ILS over MCTS}\label{sec:explaination}

Unlike MCTS, ILS does not generate a search tree and only manipulates terminal expressions. As such, no simulation is performed and the distance function guides the search well. 
Indeed, as \cref{fig:xyntia_evol_dist_success} presents, the distance follows a step-wise progression. 
Distance evolution is drastically different from \Syntia/ and enumerative MCTS (\cref{fig:syntia_dist_timeouts}). It assesses that unlike them, \Xyntia/ is guided by the distance function.
This enables \Xyntia/ to synthesize deeper expressions that would be out of reach for enumerative search. Moreover, note that \Xyntia/ globally follows a positive trend i.e. it does not unlearn previous work. Indeed, before each perturbation, the best expression found from now is restored. Thus, if iterative mutations follows a misleading path, the resulting solution is not kept and the best solution is reused to be perturbed. Keeping the current best solution is of first relevance as the search space is highly unstable and enables \Xyntia/ to be more reliable and less dependant of randomness. 

\begin{figure}[!htbp]
    \centering
     \resizebox{.9\columnwidth}{!}{%
    \input{./graphs/ils_dist_time.tex}
    }
    \caption{\Xyntia/ (\XDef/): distance evolution}
    \label{fig:xyntia_evol_dist_success}
\end{figure}

\myparagraph{Conclusion} {\it Unlike MCTS, which is  almost enumerative in code deobfuscation, ILS is well guided by the objective function and the distance evolution throughout the synthesis follows a positive trend, hence the difference in performance. Moreover, this is true as well for other S-metaheuristics, which appear to be much more suited for code deobfuscation than MCTS.}

\subsection{Limitations}
\label{sec:limitations}

Blackbox methods
rely on two main steps, sampling and learning, which both show 
weaknesses.
Indeed, \Xyntia/ and \Syntia/ randomly sample inputs to approximate the semantics of an expression. It then assumes that samples depict all behaviors of the code under analysis. If this assumption
is invalid then the learning phase will miss some behaviors, returning partial results. As such, blackbox deobfuscation is not appropriate to handle points-to functions.  

Learning can itself be impacted by other factors. For instance,  learning
expressions with unexpected constant values is hard. 
Indeed, the grammar of \Xyntia/ and \Syntia/ only considers  constant value 
\texttt{1}. Thus,  finding expressions with constant values absent from the grammar
requires to   create them (e.g., encoding 3 as $1+1+1$), which may be 
unlikely. A naive solution is to add to the grammar additional  constant values but
it  significantly impacts efficiency. \cref{annex:limitations} studies the effect of introducing higher numbers of constant values in \Xyntia/. 
For 100 values, the equivalence rate is divided by 2 (resp., by 4 for 200 values).
Still, \cref{sec:deobfuscation_xyntia} shows that \Xyntia/ can synthesize interesting constant values (unlike \Syntia/).

\subsection{Conclusion}
\label{sec:conclusion}

Because of the high instability of the search space, \textit{Iterated Local Search} is much more appropriate than MCTS (and, to a lesser extent, than other S-metaheuristics) for blackbox code deobfuscation, as it manipulates terminal expressions only and is able to restore the best solution seen so far in case the search gets lost.  These features enable \Xyntia/ to keep the advantages of \Syntia/ (stability, output quality) 
while clearly improving over its weaknesses: especially \Xyntia/ manages with 1s timeout  to synthesize twice more expressions than \Syntia/ with 10min timeout. 

Other S-metaheuristics also perform significantly better than MCTS here, demonstrating that the problem itself is not well-suited for partial solution exploration and simulation-guided search.

\begin{figure*}[!htb]
    \begin{subfigure}[t]{\columnwidth}
    \centering
     \resizebox{0.8\columnwidth}{!}{%
    \input{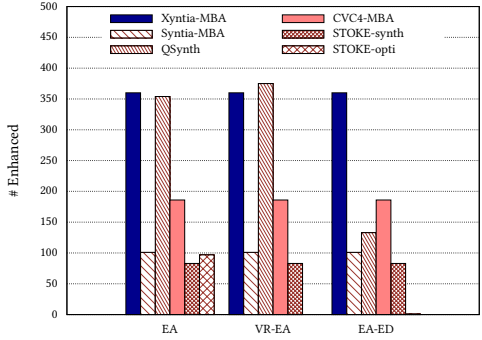}
    }
    \captionsetup{width=.9\columnwidth}
    \caption{Enhancement rate}
    \label{fig:bqs}
  \end{subfigure}
  \begin{subfigure}[t]{\columnwidth}
    \centering
     \resizebox{0.8\columnwidth}{!}{%
    \input{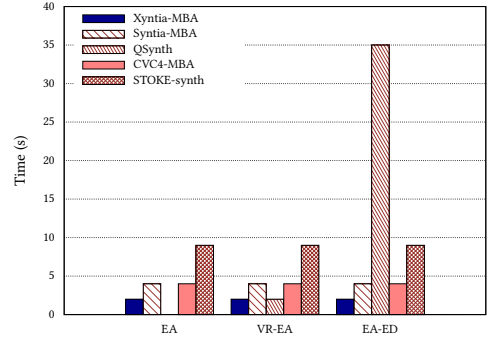}
    }
    \captionsetup{width=.9\columnwidth}
    \caption{Mean synthesis time per expression -- \stokeopti/  not
      shown as it always uses  60 s}
    \label{fig:speed_bqs}
  \end{subfigure}
  \caption{\Syntia/, \QSynth/, \Xyntia/, CVC4 and STOKE on EA, VR-EA and EA-ED datasets (timeout = 60 s)}
\end{figure*}

\section{Compare to other approaches} \label{sec:comparative_study}

We now  extend the comparison to other state-of-the-art
tools:
\begin{enumerate*}
    \item a greybox deobfuscator (\QSynth/~\cite{davidqsynth});
    \item whitebox simplifiers (GCC, Z3 simplifier and our custom simplifier);
    \item program synthesizers (CVC4~\cite{BCD+11}, winner of
      the SyGus'19 syntax-guided synthesis competition~\cite{DBLP:journals/corr/abs-1904-07146} 
      and STOKE~\cite{DBLP:journals/corr/abs-1211-0557},  an efficient superoptimizer).
\end{enumerate*}
Unlike blackbox approaches, greybox and whitebox methods  
should be evaluated on the
enhancement rate. Indeed, these methods
can always succeed by returning the obfuscated expression without 
simplification. The enhancement rate measures how often
synthesized expressions are smaller than the original ($quality \leq 1$). 

\subsection{Experimental design} 
\label{sec:experimental-design}

\Xyntia/ and \QSynth/ learn expressions over distinct grammars: \EXPR/ and \MBA/ respectively. 
Moreover, \QSynth/ is unfortunately not available, whether in a source or executable form. So we could neither adapt nor reproduce the experiments. In the end, we could
only compare it over \MBA/, using the results reported by David et
al.~\cite{davidqsynth}.

\myparagraph{Benchmarks} We compare blackbox program synthesizers on B2 and grey/white box approaches on \QSynth/'s datasets -- available for extended comparison\footnote{https://github.com/werew/qsynth-artifacts}. Thus, we consider the 3  datasets from David et al.'s \cite{davidqsynth} of obfuscated expressions using Tigress~\cite{Tigress}: 
{\bf EA} (base dataset, obfuscated with the
          \textit{EncodeArithmetic} transformation),
{\bf VR-EA} (EA obfuscated with \textit{Virtualize} and \textit{EncodeArithmetic} protections), and
	{\bf EA-ED} (EA obfuscated with \textit{EncodeArithmetic} and \textit{EncodeData} transformations.

\subsection{Comparative evaluation}

\myparagraph{Greybox} We compare \Xyntia/ to \QSynth/'s published results
 ~\cite{davidqsynth} on EA,
 VR-EA and EA-ED. 
 \cref{fig:bqs} shows  that while both tools  reach comparable results (enhancement rate $\approx$ 350/500) for simple obfuscations (EA and VR-EA),   
\Xyntia/ keeps the same results for heavy obfuscations (EA-ED) 
while \QSynth/ drops to 133/500. Actually, \Xyntia/ is insensitive to syntactic complexity while \QSynth/ is. 

\myparagraph{Whitebox} We compare \Xyntia/ over the EA, VR-EA and EA-ED datasets with 3 whitebox approaches: GCC, Z3 simplifier (v4.8.7) and our custom simplifier. As expected, they are not efficient compared to \Xyntia/ (\cref{annex:vswhite}). Regardless of the obfuscation, they simplify $\leq 68$ expressions where \Xyntia/ simplifies 360 of them.

\myparagraph{Program synthesizers} 
We now compare \Xyntia/ to state-of-the-art program
synthesizers, namely CVC4~\cite{BCD+11} and
STOKE~\cite{DBLP:journals/corr/abs-1211-0557}. 
CVC4 takes as input a grammar and a specification 
and returns, through enumerative search, a 
consistent expression. STOKE is a super-optimizer leveraging program
synthesis (based on Metropolis Hasting) to infer optimized code snippets. It does not return an
expression but optimized assembly code. STOKE addresses the
optimization problem in two ways:
\begin{enumerate*}
\item
  \stokesynth/ starts from a pre-defined number of nops and mutates them.
    \item \stokeopti/ starts from the non-optimized code and mutates it to simplify it.
\end{enumerate*}
While STOKE integrates its own sampling strategy and grammar, CVC4 does not -- thus, we consider for CVC4 the same sampling strategy as \Xyntia/ (100 I/O samples with 5 constant vectors) as well as the \EXPR/ and \MBA/ grammars. More precisely, CVC4-\EXPR/ is used over B2 to compare to \Xyntia/ (\ref{config}) and CVC4-\MBA/ is evaluated on EA, VR-EA and EA-ED to compare against \QSynth/.

\begin{table}[htbp]
	\centering
        \caption{Program synthesizers on  B2}
        \begin{adjustbox}{max width=\columnwidth}
            \begin{tabular}{lccc}
                    & CVC4-\EXPR/ & \stokesynth/ \\
                    \cmidrule(r){1-1}\cmidrule(r){2-2}\cmidrule(lr){3-3}
                    Success Rate & 36.8\% & 38.0\% \\
                    Equiv. Range & 29.3 - 36.8\% & 38.0\% \\
                    Mean Qual. & 0.56 & 0.91 \\
		\end{tabular}
	\end{adjustbox}
        \label{tab:synthesizers_b2}
\end{table}

\cref{tab:synthesizers_b2} shows that CVC4-\EXPR/ and \stokesynth/ fail to synthesize more than 40\% of B2 while \Xyntia/ reaches
90.6\% proven equivalence rate. Indeed enumerative search (CVC4) is
less appropriate when time is limited. Results
of \stokesynth/ are also expected as its search space considers all assembly mnemonics. Moreover, \cref{fig:bqs} shows that blackbox and whitebox (\stokeopti/) synthesizers do not efficiently simplify obfuscated expressions. \stokeopti/ finds 
only 1 / 500 expressions over EA-ED and does not handle jump instructions, inserted by the VM, failing to analyze VR-EA.

\subsection{Conclusion} 
\label{sec:conclusion-1}

\Xyntia/ rivals \QSynth/ on light / mild protections and outperform it on heavy protections,  
while pure whitebox approaches are far behind, showing the benefits of being independent from syntactic complexity. 
Also, \Xyntia/ outperforms state-of-the-art program synthesizers showing that it is better suited to perform deobfuscation. These good results show that seeing  deobfuscation as  
an optimization problem is fruitful.

\section{Deobfuscation with \Xyntia/} \label{sec:deobfuscation_xyntia}

We now prove that \Xyntia/ is insensitive to common protections (opaque predicates) as well as to recent anti-analysis protections (MBA, covert channels, path explosion)  
and we confirm that blackbox methods can help reverse state-of-the-art virtualization \cite{Tigress,VMProtect}. 

\subsection{Effectiveness against usual protections} \label{sec:usual_obf}

\begin{table*}[ht]
	\centering
        \caption{\Xyntia/ (\ref{config}) against usual protections (B2, timeout = 60 s)}

        \resizebox{0.7\textwidth}{!}{%
        	\begin{tabular}{cccccc}
        	%\midrule
                	& $\emptyset$ & MBA & Opaque & Path oriented & Covert channels \\
                        \cmidrule(lr){2-2}\cmidrule(lr){3-3}\cmidrule(lr){4-4}\cmidrule(lr){5-5}\cmidrule(lr){6-6}
                    
                    Succ. Rate & 95.5\% & 95.4\% & 94.68\% & 95.4\% & 95.1\% \\
                    Equiv. Range & 90.6 - 94.2\% & 90.0 - 93.8\% & 89.9 - 93.0\% & 89.5- 93.7\% & 89.0 - 94.0\% \\
                    Mean Qual. & 0.92 & 0.95 & 0.90 & 0.94 & 0.89 \\
                    \midrule    
		\end{tabular}
	}
	\label{tab:xyntia_vs_usual_protect}
\end{table*}

\Xyntia/ is able to bypass many protections (\cref{tab:xyntia_vs_usual_protect}).

\textbf{Mixed Boolean-Arithmetic}~\cite{Zhou:2007:IHS:1784964.1784971} hides the
original semantics of an expression both to humans and SMT solvers. However, the encoded expression remains equivalent to the
original one. As such, 
the semantic complexity stays unchanged, and \Xyntia/ should not be impacted.
Launching \Xyntia/ on B2 obfuscated with Tigress~\cite{Tigress} \textit{Encode Arithmetic} transformation (size of expression: x800)  
confirms that it has no impact.

\textbf{Opaque predicates}~\cite{collberg1998manufacturing}  obfuscate control
flow by creating artificial conditions in programs.  
The conditions are
traditionally tautologies 
and dynamic runs of the code will follow a unique path.
Thus, sampling is not affected and synthesis not impacted. We show it by launching \Xyntia/ over B2 obfuscated with Tigress \textit{AddOpaque}
transformation.

\textbf{Path-based obfuscation}~\cite{10.1145/2810103.2813663,ollivier2019kill}
takes advantage of the path explosion problem to thwart symbolic execution, 
massively adding additional feasible paths through dedicated encodings. 
We show that it has no effect, by protecting B2 with a custom encoding inspired by \cite{ollivier2019kill} (\cref{annex:usual_obf} gives an example of our encoding). 

\textbf{Covert channels} \cite{collbergprobabilistic} hide information flow to static analyzers by rerouting data to invisible states (usually OS related) before retrieving it -- for example taking advantage of timing difference between a slow thread and a fast thread to infer the result of some computation with great accuracy.  
Again, as blackbox deobfuscation focuses only on input-output relationship,  covert channels should  not disturb it. 
Note that the probabilistic nature of such  obfuscations (obfuscated behaviours can differ from unobfuscated ones from time to time) could be a problem in case of high fault probabilities, but in order for the technique to be useful, fault probability must precisely remains low. 
We obfuscate B2 with the
\textit{InitEntropy} and \textit{InitImplicitFlow} (thread kind) transformations
of Tigress \cite{Tigress}.  \cref{tab:xyntia_vs_usual_protect} indeed shows the absence
of impact: ``faults'' probability being so low, it does not
affect sampling.

\myparagraph{Conclusion} \textit{State-of-the-art protections are not effective against blackbox deobfuscation. They prevent efficient reading 
of the code and tracing of data but blackbox methods directly execute it.}

\subsection{Virtualization-based obfuscation}\label{sec:usecase}

We now use \Xyntia/ to reverse code obfuscated with state-of-the-art virtualization. We obfuscate a program computing MBA operations with Tigress \cite{Tigress}
and VMProtect \cite{VMProtect} and  
our goal is to reverse the VM handlers. Using such synthetic program enables to expose a wide variety of handlers. 

\begin{table}[ht]
 \caption{\Xyntia/ and \Syntia/ results over program obfuscated with Tigress \cite{Tigress} and VMProtect \cite{VMProtect}}
	\centering
       
        \begin{adjustbox}{max width=\columnwidth}
            \begin{tabular}{llccc}
            & & Tigress (simple) & Tigress (hard) & VMProtect \\
           \cmidrule(lr){2-5}
            & Binary size & 40KB & 251KB & 615KB \\
            & \# handlers & 13 & 17 & 114 \\
            & \# instructions per handlers & 16 & 54 & 43 \\
            \midrule
            \multirow{2}{*}{\Xyntia/} & Completely retrieved  & 12/13 & 16/17 & 0/114 \\
            & Partially retrieved  & 13/13 & 17/17 & 76/114 \\
            \midrule
             \multirow{2}{*}{\Syntia/} & Completely retrieved  & 0/13 & 0/17 & 0/114 \\
            & Partially retrieved  & 13/13 & 17/17 & 76/114 \\
		\end{tabular}
	\end{adjustbox}
	
	\label{tab:vmobfstats}
\end{table}

\noindent\textbf{Tigress}~\cite{Tigress} is a source-to-source obfuscator. 
Our obfuscated program contains 13 handlers. 
Since at assembly level  each handler ends with an indirect jump to the next handler to execute,  we were able to extract the positions of handlers using execution traces. We then used the scripts from \cite{Blazytko:2017:SSS:3241189.3241240} to sample each handler. 
\Xyntia/ synthesizes 12/13 handlers in less that 7~s each. We can classify  them in different categories: 
\begin{enumerate*}
    \item arithmetic and Boolean ($+$, $-$, $\times$, $\land$, $\lor$, $\oplus$);
    \item stack (store and load); 
    \item control flow (goto and return);
    \item calling convention 
     (retrieve obfuscated function's arguments). 
\end{enumerate*}
These results show that \Xyntia/ can synthesize a wide variety of handlers. 
Interestingly,  while these handlers contain many constant values (typically, offsets for context update), 
\Xyntia/ can handle them as well. 
In particular, it infers the calling convention related handler, synthesizing constant values up to 28 (to access the 6th argument).
Thus, even if \Xyntia/ is inherently limited on constant values (see \cref{sec:limitations}) it still handles them to a limited extent.  
Repeating the experiment by  adding \textit{Encode Data} and \textit{Encode Arithmetic}  to \textit{Virtualize} yields similar results. \Xyntia/ synthesizes all 17 exposed handlers but 
one,  
 confirming that \Xyntia/ handles combinations of protections. 
 Finally, note that \Syntia/ fails to synthesize handlers completely (not handling constant values). Still it infers arithmetic and Boolean handlers (without context updates).  

\smallskip 

\noindent\textbf{VMProtect} \cite{VMProtect} is an assembly to assembly obfuscator. We use the latest premium version (v3.5.0). As each VM handler ends with a \lstinline{ret} or an indirect jump,  we easily extracted each distinct handler from execution traces.  Our traces expose 114 distinct handlers containing on average 43 instructions (\cref{tab:vmobfstats}). VMProtect's VM is stack based.  
To infer the semantics of each handler, we again used Blazytko's scripts~\cite{Blazytko:2017:SSS:3241189.3241240} in ``memory mode'' (i.e., forbidding registers to be seen as inputs or outputs). Our experiments show that each arithmetic and Boolean handlers (\lstinline{add}, \lstinline{mul}, \lstinline{nor}, \lstinline{nand}) are replicated 11 times to fake a large number of distinct handlers. Moreover, we are also able to extract the semantics of some stack related handlers. In the end, we successfully infer the semantics of 44 arithmetic or Boolean handlers and 32 stack related handlers. Synthesis took at most 0.3~s per handler. 
\Syntia/ gets equal results as \Xyntia/. 

\myparagraph{Conclusion} \textit{\Xyntia/ synthesizes most Tigress' VM handlers, (inclduing interesting constant values)  and extracts the semantics of VMProtect's arithmetic and Boolean handlers. This shows that blackbox deobfuscation can be highly effective, making the need for efficient protections clear.} 

\section{Counter AI-based Deobfuscation} \label{sec:protections}

We now study defense mechanisms against blackbox deobfuscation. 

\subsection{General methodology}

We remind that blackbox methods require  the reverser to locate
a suitable reverse window delimiting the code of interest with its input and
output. This can be done manually or automatically~\cite{Blazytko:2017:SSS:3241189.3241240}, still this is mandatory  and not trivial.  The defender could target this step, reusing standard obfuscation techniques. 

{\it Still there is a risk that the attacker finds the good windows. Hence we are looking for a more radical protection against blackbox attacks. We suppose that the reverse window,  input and output are correctly identified, and we seek to protected a given piece of code. 
}

Note that adding extra {\it fake}  inputs (not influencing the result) is easily circumvented in a blackbox setting, 
by dynamically testing different values for each input and filtering inputs where no difference is observed.  

\myparagraph{Protection rationale} Even with correctly delimited windows, synthesis
can still be thwarted.  Recall that blackbox methods rely on 2 main steps
\begin{enumerate*}
    \item I/O sampling; 
    \item learning from samples,
\end{enumerate*}
and both can be sabotaged. 

\begin{itemize}
\item First, if the sampling phase is not performed properly,
the learner could miss important behaviors of the code, returning incomplete or even misleading information;  

\item Second, if the expression under analysis is too complex, the learner will fail to map inputs to their outputs.
\end{itemize}

In both cases, no information is retrieved. 
Hence, the key to impede blackbox deobfuscation is to migrate \textit{from syntactic complexity  to semantic complexity}. 
 We propose in \cref{sec:semant_complex,sec:merged_handlers}  
two novel protections impeding the sampling and learning phases.

\subsection{Semantically complex handlers} \label{sec:semant_complex}

Blackbox approaches are sensitive to semantic complexity. As such,
relying on a set of complex handlers is an effective strategy to thwart
synthesis. These complex handlers can then be combined to recover standard operations. 
We propose a method to generate arbitrary complex
handlers in terms of size and number of inputs.

\myparagraph{Complex semantic handlers} Let $S$ be a set of expressions and $h, e_1, ..., e_{n-1}$ be $n$ expressions in $S$.
Suppose that $(S, \star)$ is a group. Then $h$ can be encoded as $h = \underset{i=0}{\overset{n-1}{\star}} h_i$, where for all i, with $0 \leq i \leq n$,
\[
h_i = 
\left \{
    \begin{array}{l l}
        h - e_1  & \textrm{if } i = 0 \\
        e_i - e_{i+1} & \textrm{if } 1 \leq i < n-1 \\
	e_{n-1}   & \textrm{if } i = n-1\\
    \end{array}
\right .
\]

\noindent Each $h_i$ is a new handler that can be combined with others to express common operations -- see \cref{tab:encoding_ex} for an example. Note that the choice of $(e_1, ..., e_n)$ is arbitrary. 
One can choose very complex
expressions with as many arguments as wanted.

\begin{table}[ht]
	\centering
        \caption{Examples of encoding}
        \begin{adjustbox}{max width=\columnwidth}
        	\begin{tabular}{rc@{ = }c}
		  & $h_0$ & $(x + y) + -((a -x^2) - (xy))$\\
		+ & $h_1$ & $(a -x^2) - xy + (- (y - (a \land x)) \times (y \otimes x))$\\
		+ & $h_2$ & $(y - (a \land x)) \times (y \otimes x)$\\ \hline
		  & $h$   & $x + y$\\
		\end{tabular}
	\end{adjustbox}
	\label{tab:encoding_ex}
\end{table}
\myparagraph{Experimental design} To evaluate \Syntia/ and \Xyntia/ against our new encoding, we created 3 datasets -- BP1, BP2 and BP3, listed by increasing order of complexity.
Each dataset contains 15 handlers which can be combined to encode the $+, - , \times, \land$ and  $\lor$ operators.
Within dataset, all handlers have the same number of inputs. \cref{tab:bpstats} reports details on each datasets -- more details are available in \cref{annex:protections}. The mean overhead column is an estimation of the complexity added to the code by averaging the number of operators needed to encode a single basic operator ($+, -, \times,\vee,\wedge$). Overheads in BP1 (21x), BP2 (39x) and even BP3 (258x)   are reasonable compared to some syntactical obfuscations: encoding $x+y$ with MBA three times in Tigress yields a 800x overhead. 

\myparagraph{Evaluation} Results (\cref{fig:bps_results}) show that while \Xyntia/ (with 1h.expr.) manages well low complexity handlers (BP1: 13/15), yet performance degrades quickly as complexity increases (BP2: 3/15, BP3: 1/15).  \Syntia/, CVC4 and \stokesynth/ find none with 1~h/expr., even on BP1 (\cref{annex:protections}).

\begin{table}[ht]
	\centering
        \caption{Protected datasets}
        \begin{adjustbox}{max width=\columnwidth}
            \begin{tabular}{lcccccc}
                & \#exprs & min size & max size & mean size & \#inputs & mean overhead \\
                    \cmidrule(lr){1-1}\cmidrule(lr){2-2}\cmidrule(lr){3-3}\cmidrule(lr){4-4}\cmidrule(lr){5-5}\cmidrule(lr){6-6}\cmidrule(lr){7-7}
		BP1 & 15 & 4 & 11 & 6.87 & 3 & x21 \\
		BP2 & 15 & 8 & 21 & 12.87 & 6 & x39 \\
		BP3 & 15 & 58 & 142 & 86.07 & 6 & x258 \\
		\end{tabular}
	\end{adjustbox}
	\label{tab:bpstats}
\end{table}

\myparagraph{Conclusion} {\it Semantically complex handlers are efficient against  blackbox  deobfuscation. While high complexity handlers comes with a cost similar to strong MBA encodings, medium complexity handlers offer a strong protection at a reasonable  cost.  
 }
 
\myparagraph{Discussion} Our protection can be bypassed if the attacker focuses on the good combinations of handlers, rather than on the handlers themselves. To prevent it, complex handlers can be duplicated (as in VMProtect, see \cref{sec:usecase}) to make patterns recognition more challenging.

\begin{figure}[!ht]
    \centering
     \resizebox{0.8\columnwidth}{!}{%
        \input{./graphs/xyntia_bps.tex}
      }
    \caption{\Xyntia/ (\ref{config}) on BP1,2, 3 -- varying timeouts}
    \label{fig:bps_results}
\end{figure}

\subsection{Merged handlers} \label{sec:merged_handlers}

We now propose another protection, based on conditional expressions and the merging of existing handlers. While block merging is known for a long time against human reversers, we show that it is extremely efficient against blackbox attacks.  
Note that while we write our merged  handlers  with explicit if-then-else operators (ITE) for simplicity,  these conditions are not necessarily 
implemented  with conditional branching  (cf.~\cref{lst:branch-branchless} for an example of branchless encoding). 
Hence, we consider that the attacker sees merged handlers as a unique code fragment.

\myparagraph{Datasets} We introduce 5 datasets\footnote{Available at : \datasetUrl{}} (see \cref{annex:merged}) composed of  20 expressions. Expressions in dataset 1  are built with  1 \textit{if-then-else} (ITE) exposing 2 basic handlers (among $+, -, \times, \land, \lor, \oplus$); expressions in dataset 2 are built with 2 nested ITEs exposing 3 basic handlers, etc.  Conditions are equality  checks against consecutive constant values ($0, 1, 2$, etc.).  
For example, dataset 2 contains the expression:
\begin{equation} \label{eq:example_ite}
    ITE(z=0, x+y, ITE(z=1, x-y, x \times y))
\end{equation}

\myparagraph{Scenarios} Adding conditionals brings extra challenges 
\begin{enumerate*}
    \item the grammar must be expressive enough to handle conditions;
    \item the sampling phase must be efficient enough to cover all possible behaviors.
\end{enumerate*}
Thus, we consider different scenarios: 
\begin{itemize}[leftmargin=* ,parsep=0cm,itemsep=0cm,topsep=0cm, align=left]
\item[\textit{Utopian} \hspace{1pt}] The synthesizer learns expressions over the
  \MBA/ set of operators, extended with an $ITE(\star = 0, \star, \star)$
  operator (\MBA/+ITE operator set). Moreover, the sampling is done so that all branches are traversed the same
  number of time. This situation, favoring the attacker, will show that
  merged handlers are always efficient.
    \item[\textit{\MBA/ + ITE} \hspace{1pt}] This situation is more realistic: the attacker does not know at first glance how to sample. However, its grammar fits perfectly the expressions to reverse.
    \item[\textit{\MBA/ + Shifts} \hspace{1pt}] Here \Xyntia/ does not sample
      inputs uniformly over the different behaviors, does not consider ITE
      operators, but allows shifts to represent branch-less conditions.
    
    \item[\textit{Default.} \hspace{1pt}] This is the default version of the synthesizer.  
\end{itemize}

In all these scenarios, appropriate constant values are added to the grammar. For example, to synthesize \cref{eq:example_ite}, 
\texttt{0} and \texttt{1} are added.  

\begin{figure}[!ht]
      \begin{lstlisting}[numbers=none]
int32_t h(int32_t a, int32_t b, int32_t c) {
    // if (c == cst) then h1(a,b,c) else h2(a,b,c);
    int32_t res = c - cst ;
    int32_t s = res >> 31;
    res = (-((res ^ s) -s) >> 31) & 1;
    return h1(a, b, c)*(1 - res) + res*h2(a, b, c);
}\end{lstlisting}
  
  \caption{Example of a branch-less condition}
    \label{lst:branch-branchless}
\end{figure}

\begin{figure}[!ht]
\centering
\resizebox{0.8\columnwidth}{!}{%
        \input{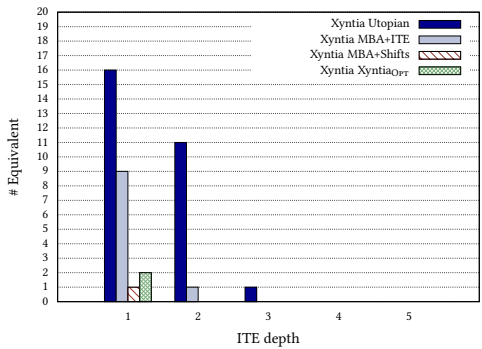}
    }
\caption{Merged handlers: \Xyntia/ (timeout=60s)}
\label{fig:merged}
\end{figure}

\myparagraph{Evaluation} \cref{fig:merged} presents \Xyntia/'s results  on the 5
datasets. As expected, the \textit{Utopian} scenario is where \Xyntia/ does best, still it cannot cope with more than 3 nested ITEs. For realistic scenarios, \Xyntia/ suffers even more. 
Results for \Syntia/, CVC4 and \stokesynth/ 
(see \cref{annex:merged}) confirm this result (no solution found for $\geq 2$ nested ITEs). 
Note that overhead here is minimal, as the merging only add 
a few conditional jumps. 

\myparagraph{Conclusion} {\it Merged handlers are extremely powerful against  blackbox synthesis. 
Even in the ideal sampling scenario, 
blackbox methods
cannot retrieve the semantics of expressions with more than 3 nested conditionals -- while runtime overhead is minimal.}

\myparagraph{Discussion}
Symbolic methods, like symbolic execution, are unhindered by this protection, for they track the succession of handlers and know which sub parts of merged
handlers are executed. They can then reconstruct the real semantics of the
code. To handle this, our anti-AI protection can be combined with (lightweight) 
anti-symbolic protections (e.g.~\cite{10.1145/2810103.2813663,ollivier2019kill}).

\section{Related Work}

\myparagraph{Blackbox deobfuscation} Blazytko et al.'s work~\cite{Blazytko:2017:SSS:3241189.3241240}
has already been thoroughly discussed. 
We complete their experimental evaluation, generalize and improve their approach: 
\Xyntia/ with 1s/expr. finds twice more expressions than \Syntia/ with 600s/expr.

\myparagraph{White- and greybox deobfuscation} Several recent works leverage  {\it whitebox}  symbolic methods
for deobfuscation (``symbolic deobfuscation'') \cite{Schrittwieser:2016,DBLP:conf/sp/BardinDM17,YadegariJWD15,DBLP:series/ais/BrumleyHLNSY08,SalwanBarPot18,DBLP:conf/wcre/Kinder12}. Unfortunately, they are sensitive to code complexity as discussed in \cref{sec:deobfuscation_xyntia}, and efficient countermeasures are now available \cite{ollivier2019kill, ollivier2019obfuscation,Zhou:2007:IHS:1784964.1784971,Collberg:2009:SSO:1594894} -- while \Xyntia/ is immune to them (\cref{sec:usual_obf}).  
David et al. \cite{davidqsynth} recently proposed QSynth, 
a {\it greybox} deobfuscation method combining  I/O relationship caching (blackbox) and incremental reasoning along the target expression (whitebox).  
 Yet, \QSynth/ is sensitive to massive syntactic obfuscations where \Xyntia/ is not (cf.~\cref{sec:comparative_study}). 
Furthermore, QSynth works on a simple grammar. 
It is unclear whether its caching technique  would scale to larger grammars like those of \Xyntia/ and \Syntia/. 

\myparagraph{Program synthesis} Program synthesis aims at finding a function
from a specification which can be given either formally, 
in natural language or {\it as I/O relations} -- the case we are interested in here. 
There exist three main families of program synthesis methods \cite{gulwani2017program}: enumerative, constraint solving and stochastic. 
Enumerative search 
does enumerate all programs starting from the simpler one, pruning snippets incoherent with the specification and returning the first code meeting the specification. We compare, in this paper, to one of such method -- CVC4 \cite{BCD+11}, winner of the SyGus '19 syntax-guided synthesis competition \cite{DBLP:journals/corr/abs-1904-07146} -- and showed that our approach is more appropriate to deobfuscation. 
Constraint 
solving methods \cite{jha2010oracle} on the other hand encode the skeleton of the target program
as a first order satisfiability problem and use an off-the-shelf SMT solver to infer an
implementation meeting specification. 
However, it is less efficient than enumerative and stochastic methods \cite{DBLP:conf/fmcad/AlurBJMRSSSTU13}.  
Finally, stochastic methods \cite{DBLP:journals/corr/abs-1211-0557}  traverse the search space randomly in the hope of finding a program consistent with a  specification. 
Contrary to them,  
we aim at solving the deobfuscation problem in a {\it fully} blackbox way (not relying on the obfuscated code, nor on an estimation of the result size). 

\section{Conclusion}

AI-based blackbox deobfuscation is a promising recent research area. The field 
has been barely explored yet and the pros and cons of such methods are still unclear. This article deepens the state of AI-based blackbox deobfuscation in three different directions. 
First, we define a novel generic 
framework for AI-based blackbox deobfuscation, encompassing  prior works such as \Syntia/, we identify  that the search space underlying code deobfuscation is too unstable for simulation-based methods, and advocate the use of S-metaheuritics. 
Second we take advantage of our framework to carefully design \Xyntia/, a new  AI-based blackbox deobfuscator. \Xyntia/ significantly outperforms \Syntia/ in terms of success rate, while keeping its good properties -- especially, \Xyntia/ is completely immune to the most recent anti-analysis code obfuscation methods. 
\Xyntia/ also proves to be more efficient  than greybox and whitebox deobfuscators or standard program synthesis methods.  
Finally, we propose the two first  protections against AI-based blackbox deobfuscation, completely preventing   \Xyntia/ and \Syntia/'s attacks for reasonable cost.  
We hope that these results will help better understand AI-based
deobfuscation, and lead to further progress in the field.

\newpage
\bibliographystyle{ACM-Reference-Format}
\bibliography{sample-sigconf}

\newpage
\appendix

\section{Appendix}

We now introduce complementary results to describe details that we did not fully explained for the sake of space. We follow the same organisation as the main article:
\begin{itemize}
    \item[] {\bf \cref{annex:lessons_syntia}} details the evaluation of \Syntia/ (from \cref{sec:lesson_syntia}). It presents used datasets and obtained \Syntia/ results;
    \item[] {\bf \cref{annex:xyntia_experimental_eval}} details the evaluation of \Xyntia/ (see \cref{sec:xyntia}) and the study leading to the optimal \Xyntia/;
    \item[] {\bf \cref{annex:compare_other}} describes the comparison of \Xyntia/ to whitebox, pattern based simplifiers from \cref{sec:comparative_study}.
    \item[] {\bf \cref{annex:obf}} details the obfucations used to evaluate \Xyntia/ against state-of-the-art protections in \cref{sec:deobfuscation_xyntia};
    \item[] {\bf \cref{annex:protections}} describes datasets used in \cref{sec:protections} and details evaluation of \Syntia/, CVC4 and STOKE over proposed protections.
\end{itemize}

\subsection{Understand AI-based deobfuscation: More Details} \label{annex:lessons_syntia}

\cref{sec:lesson_syntia} presents an in-depth evaluation of \Syntia/. We show now complementary data to detail:
\begin{enumerate*}
    \item the distribution of expressions in our custom benchmark suite B2;
    \item the results of \Syntia/ over 15 runs;
    \item the results of \Syntia/ in terms of quality and correctness and how it reacts to the number of inputs and expression types;
    \item the study of \Syntia/'s parameters to find an optimal configuration.
\end{enumerate*}

\begin{table}[ht]

  \centering
  \caption{Description of B2}
  \resizebox{\columnwidth}{!}{%
  \begin{tabular}{lcccccccc}
      & \multicolumn{3}{c}{Type} & \multicolumn{5}{c}{\# Inputs} \\
      \cmidrule(r){2-4}\cmidrule(r){5-9}
      & Bool. & Arith. & MBA & 2 & 3 & 4 & 5 & 6  \\
        \midrule
      \#Expr. &  370 & 370 & 370 & 150 & 600 & 180 & 90 & 90 \\
  \end{tabular}
  }

\label{tab:b2_stats}
\end{table}

\subsubsection{\textbf{Experimental design}} In order to perform a fine grained evaluation of \Syntia/, we use 2 benchmark suites: B1 and B2. B1 has been introduced by Blazytko et al. \cite{Blazytko:2017:SSS:3241189.3241240} to evaluate \Syntia/, and contains 500 expressions. However, it presents important limitations as discussed in \cref{sec:syntia_eval}. Thus, we introduce a custom benchmark B2 which contains 1110 expressions. It is better distributed according to the type of the expressions -- Boolean, Arithmetic and Mixed Boolean-Arithmetic -- and number of inputs used -- between 2 and 6. Moreover, B2 is more challenging than B1, considering more complex expressions. \cref{tab:b2_stats} presents the number of expressions per type of expressions and number of inputs.

\subsubsection{\textbf{Evaluation of \Syntia/}} In \cref{sec:syntia_eval} we evaluate \Syntia/ to estimate its stability across executions, its robustness, speed, quality and correctness. We present now complete experiments results and discuss them.

\myparagraph{\ref{qq1}} \cref{tab:15runs} presents results of \Syntia/ over 15 runs and \cref{tab:svariability} presents statistics on it. We observe that \Syntia/ is indeed very stable across executions. 

\begin{table}[ht]
  \centering
  \caption{Success rate of \Syntia/ across 15 runs (timeout=60s)}
  \resizebox{.8\columnwidth}{!}{%
  {\footnotesize 
 \begin{tabular}{@{}ccc@{}}
    Test execution no. & B1 & B2 \\
    \cmidrule(lr){1-1} \cmidrule(lr){2-2}\cmidrule(lr){3-3}
    1 & 367 (73.4\%) & 349 (31.4\%) \\
    2 & 362 (72.4\%) & 376 (33.9\%) \\
    3 & 376 (75.2\%) & 371 (33.4\%) \\
    4 & 365 (73.0\%) & 367 (33.1\%) \\
    5 & 369 (73.8\%) & 379 (34.1\%) \\
    6 & 365 (73.0\%) & 383 (34.5\%) \\
    7 & 375 (75.0\%) & 366 (33.0\%) \\
    8 & 370 (74.0\%) & 371 (33.4\%) \\
    9 & 366 (73.2\%) & 358 (32.3\%) \\
    10 & 372 (74.4\%) & 367 (33.1\%) \\
    11 & 367 (73.4\%) & 364 (32.8\%) \\
    12 & 364 (72.8\%) & 372 (33.5\%) \\
    13 & 371 (74.2\%) & 378 (34.1\%) \\
    14 & 368 (73.6\%) & 350 (31.5\%) \\
    15 & 370 (74.0\%) & 354 (31.9\%) \\
  \end{tabular}
  }
  }
  \label{tab:15runs}
\end{table}

\begin{table}[ht]
  \centering
  \caption{15 runs of \Syntia/ over B1 and B2 (timeout = 60 s)}
  \resizebox{\columnwidth}{!}{%
 \begin{tabular}{lccccc}
      & Data-set & Min. & Max. & Mean & $\sigma$ \\
    \cmidrule(lr){1-2}\cmidrule(lr){3-3}\cmidrule(lr){4-4}\cmidrule(lr){5-5}\cmidrule(lr){6-6}
    \multirow{2}{*}{\Syntia/}& B1 & 362(72.4\%) & 376(75.2\%) & 368.5(73.7\%) & 3.83(0.76\%) \\
    & B2 & 349(31.4\%) & 383(34.5\%) & 367.0(33.1\%) & 10.11(0.91\%) \\
  \end{tabular}
  }
  \label{tab:svariability}
\end{table}

\myparagraph{\ref{qq2}} As presented in \cref{tab:svariability} \Syntia/ is not able to synthesize B2 efficiently, only synthesizing 34.5\% of it. Moreover, as presented in \cref{tab:xyntia_distances_types}, \Syntia/ cannot handle expressions using more than 3 inputs. Indeed, its success rate falls to 10.0\%, 2.2\% and 1.1\% for respectively 4, 5 and 6 inputs. \Syntia/ is also impacted by the type of the target expression. Handling boolean expressions seems simpler for \Syntia/. On the contrary, it struggles to synthesize MBA expressions. Still we observe that \Syntia/ returns really good quality results ($\approx 0.60)$ and almost never returns non equivalent expressions.

\myparagraph{\ref{qq3}} \Syntia/ defaults to synthesizing expressions over the \FULL/ operators' set. To evaluate its sensitivity to the size of the operators' set, we launch it over \FULL/, \EXPR/ and \MBA/. \cref{tab:syntia_opsets} shows that restricting the search space benefits to \Syntia/. However, even in the best scenario (\MBA/) its results are deceiving. Indeed, it synthesizes only $\approx 42\%$ of B2. 
\begin{table}[ht]
	\centering
        \caption{\Syntia/'s results on \FULL//\EXPR//\MBA/ (B2, timeout=60s).}
        \begin{adjustbox}{max width=\columnwidth}
        	\begin{tabular}{lcccc}
                	& & \FULL/ & \EXPR/ & \MBA/ \\
                        \cmidrule(r){1-2}\cmidrule(lr){3-3}\cmidrule(lr){4-4}\cmidrule(lr){5-5}%\cmidrule(lr){6-6}
                        \multirow{4}{*}{\Syntia/} & Succ. Rate & 34.5\% & 38.8\% & 42.6\% \\
			& Equiv. Range & 33.7 - 34.0\% & 38.7\% & 42.3 - 42.6\% \\
			& Mean Qual. & 0.59 & 0.62 & 0.66 \\
		\end{tabular}
	\end{adjustbox}
	\label{tab:syntia_opsets}
\end{table}

\subsubsection{\textbf{Optimal \Syntia/}} To ensure conclusions given in \cref{sec:syntia_explaination} apply to MCTS and not only to \Syntia/, we studied \Syntia/ extensively, searching for better set-ups. We study \Syntia/ according to following parameters: simulation depth, SA-UCT value, number of I/O samples and choice of the distance.

\begin{table}[ht]
	\centering
        \caption{\Syntia/ depending on max playout depth (\MBA/, B2, timeout = 60 s).}
        \begin{adjustbox}{max width=0.5\textwidth}
        	\begin{tabular}{lccc}
        	    Max play. depth & 0 & 3 & 5 \\ 
                \cmidrule(r){1-1}\cmidrule(lr){2-2}\cmidrule(lr){3-3}\cmidrule(lr){4-4}
                Succ. Rate & 42.6 \% & 31.8 \% & 28.6 \% \\
			    Equiv. Range & 42.3 - 42.6 \% & 31.4 - 31.8 \%  & 28.1 - 28.6 \% \\
			    Mean Qual. & 0.66 & 1.03 & 1.06 \\ \\
		    \end{tabular}
	    \end{adjustbox}
	\label{tab:syntia_sim_depth}
\end{table}

\myparagraph{Simulation depth} As presented in \cref{sec:syntia_explaination}, MCTS simulates each generated nodes. To do so, it applies rules of the grammar randomly to the non terminal expression until it becomes terminal. An important parameter is thus the maximum simulation depth i.e. the number of rules not leading to terminal nodes (like $U \rightarrow U+U$). By default, \Syntia/ considers a maximum simulation depth of 0, which mean that all non terminal symbols are directly replaced by variables or constant values. \cref{tab:syntia_sim_depth} shows that increasing this parameter is not beneficial. 

\myparagraph{Number of I/O samples} By defaults \Syntia/ considers 50 samples. \cref{tab:syntia_nbsamples} presents results for different number of samples. We observe little improvement when the number of samples decreases. Still, it stays in the same range of results.  

\begin{table}[!ht]
    \centering
    \caption{\Syntia/  for different number of samples (B2, \MBA/, timeout=60s).}
    \begin{adjustbox}{max width=\columnwidth}
        \begin{tabular}{lcccc}
            \# samples & 10 & 20 & 50 & 100 \\
            \cmidrule(r){1-2}\cmidrule(lr){3-3}\cmidrule(lr){4-4}\cmidrule(lr){5-5}%\cmidrule(lr){6-6}
            Succ. Rate & 45.6\% & 44.9\% & 42.6\% & 43.2\% \\
            Equiv. Range & 45.1 - 45.4\% & 44.7 - 44.9\% & 42.3\% - 42.6\% & 42.9 - 43.2\% \\
            Mean Qual. & 0.69 & 0.71 & 0.66 & 0.69 \\
        \end{tabular}
    \end{adjustbox}
    \label{tab:syntia_nbsamples}
\end{table}

\myparagraph{Objective function} By default, \Syntia/ evaluates if an expression is close to the target one by computing the mean between different distances. To complete our evaluation of \Syntia/ we launched it with \Xyntia/'s Log-arithmetic distance. We observe that as \Xyntia/ the log-arithmetic seems more appropriate to guide the search. Still, \Syntia/'s success rate stays bellow $50\%$.

\begin{table}[!ht]
    \centering
    \caption{\Syntia/ depending on the objective function (B2, \MBA/, timeout=60s).}
    \begin{adjustbox}{max width=\columnwidth}
        \begin{tabular}{lcc}
            & \Syntia/-dist & Log-arith \\
            \cmidrule(r){1-2}\cmidrule(lr){3-3}
            Succ. Rate & 42.6\% & 47.9\% \\
            Equiv. Range & 42.3 - 42.6\% & 47.4 - 47.9\% \\
            Mean Qual. & 0.66 & 0.70 \\
        \end{tabular}
    \end{adjustbox}
    \label{tab:syntia_dist}
\end{table}

\myparagraph{Simulated annealing UCT (SA-UCT)} From a high level, MCTS can be divided in 2 behaviors: exploitation (where it focuses on promising nodes) and exploration (where it checks rarely visited or at first glance non interesting nodes). The SA-UCT constant is a parameter to configure the balance between these behaviors. The smaller is the constant the more exploitative MCTS is. On the contrary, the bigger it is, more explorative is MCTS. By default \Syntia/ sets the SA-UCT constant to 1.5. \cref{tab:syntia_sa-uct} presents results of \Syntia/ for smaller and bigger values. For smaller values, \Syntia/ is less efficient. This is coherent with claims from \cref{sec:syntia_explaination}. Indeed, as the search space is highly unstable, simulations are misleading. Thus, focusing too much on exploitation is unsuitable. However, it also appears that, bigger values can be beneficial. This is also coherent with \cref{sec:syntia_explaination} as it shows that the most important behavior is exploration. Still, even with SA-UCT values $>1.5$ success rate stays low ($<50\%$).

\begin{table}[ht]
	\centering
        \caption{\Syntia/ depending on SA-UCT value (\MBA/, B2, timeout = 60 s).}
        \begin{adjustbox}{max width=0.5\textwidth}
        	\begin{tabular}{lccccc}
        	    SA-UCT & 3 & 2 & 1.5 & 0.5 & 0.1 \\ 
                \cmidrule(r){1-1}\cmidrule(lr){2-2}\cmidrule(lr){3-3}\cmidrule(lr){4-4}\cmidrule(lr){5-5}\cmidrule(lr){6-6}
                Succ. Rate & 48.0\% & 48.2\% & 42.6 \% & 34.6 \% & 19.1 \% \\
			    Equiv. Range & 47.7 - 48.0\% & 48.1 - 48.2 \% & 42.3 - 42.6 \% & 34.6 \%  & 19.1 \% \\
			    Mean Qual. & 0.71 & 0.72 & 0.66 & 0.62 & 0.44 \\ \\
		    \end{tabular}
	    \end{adjustbox}
	\label{tab:syntia_sa-uct}
\end{table}

\myparagraph{Optimal \Syntia/} Our extensive study highlights a new optimal configuration of \Syntia/ (\MBA/ set of operators, simulation depth=0, \#samples=10, objective function=log-arithmetic, SA-UCT=2). However, even with this configuration, \Syntia/ success rate stays around 50\% (\cref{tab:syntia_opti}). While slightly better, such results are still disappointing.   
\begin{table}[ht]
	\centering
        \caption{Optimal \Syntia/ (B2, timeout = 60 s).}
        \begin{adjustbox}{max width=0.5\textwidth}
        	\begin{tabular}{lc}
        	    Succ. Rate & 52.7\% \\
        	    Equiv. Range & 52.1 - 52.6\%\\
        	    Mean Qual. & 0.76 \\
		    \end{tabular}
	    \end{adjustbox}
	\label{tab:syntia_opti}
\end{table}

\subsection{Improve AI-based deobfuscation : More details} \label{annex:xyntia_experimental_eval}

\cref{sec:xyntia} presents our new AI-based blackbox deobfuscator dubbed \Xyntia/. We show now complementary data and results to detail:
\begin{enumerate*}
    \item the results of \Xyntia/ over 15 runs;
    \item the results of \Xyntia/ in terms of quality, correctness, capacity to handle high number of inputs and different expression types;
    \item the results of \Xyntia/ over \FULL/, \EXPR/, \MBA/;
    \item the study leading to optimal \Xyntia/;
    \item the capacity of \Xyntia/ to integrate a high number of constant values in its grammar.
\end{enumerate*}

\subsubsection{\textbf{Evaluation of \Xyntia/}} To evaluate \Xyntia/ and compare it against \Syntia/ we replicate for \Xyntia/ the experimental procedure followed in \cref{sec:lesson_syntia}. We present now complete experiments results and discuss them. 

\myparagraph{\ref{q1}} To assess the usability of \Xyntia/ we need to know if it is 
stable across executions. Indeed, \Xyntia/, as \Syntia/, is stochastic and results may vary from one run to another. \cref{tab:x15runs} shows results of \Xyntia/ over 15 runs on B1 and B2 -- statistics are given in \cref{tab:xvariability}. No significant variation is observed, meaning that \Xyntia/ is stable across executions.

\begin{table}[ht]
  \centering
  \caption{Success rate of \Xyntia/ (\ref{config}) across 15 runs (timeout = 60 s)}
  \resizebox{.8\columnwidth}{!}{%
  {\footnotesize 
 \begin{tabular}{@{}ccc@{}}
      Test execution no. & B1 & B2 \\
    \cmidrule(lr){1-1} \cmidrule(lr){2-2}\cmidrule(lr){3-3}
    1 & 500 (100\%) & 1051 (94.7\%)  \\
    2 & 500 (100\%) & 1051 (94.7\%)  \\
    3 & 500 (100\%) & 1060 (95.5\%) \\
    4 & 500 (100\%) & 1054 (95.0\%) \\
    5 & 500 (100\%) & 1060 (95.5\%) \\
    6 & 500 (100\%) & 1059 (95.4\%) \\
    7 & 500 (100\%) & 1051 (94.7\%)\\
    8 & 500 (100\%) & 1059 (95.4\%)\\
    9 & 500 (100\%) & 1055 (95.0\%)\\
    10 & 500 (100\%) & 1053 (94.7\%)\\
    11 & 500 (100\%) & 1059 (95.4\%)\\
    12 & 500 (100\%) & 1052 (94.8\%)\\
    13 & 500 (100\%) & 1061 (95.6\%)\\
    14 & 500 (100\%) & 1054 (95.0\%)\\
    15 & 500 (100\%) & 1053 (94.9\%)\\
  \end{tabular}
  }
  }
  \label{tab:x15runs}
\end{table}

\begin{table}[ht]
  \centering
  \caption{\Xyntia/ (\ref{config}): 15 runs on B1/B2 (timeout=60s)}
  \resizebox{\columnwidth}{!}{%
 \begin{tabular}{lccccc}
      & Data-set & Min. & Max. & Mean & $\sigma$ \\
       \cmidrule(lr){1-2}\cmidrule(lr){3-3}\cmidrule(lr){4-4}\cmidrule(lr){5-5}\cmidrule(lr){6-6}
    \multirow{2}{*}{\Xyntia/}& B1 & 500(100\%) & 500(100\%) & 500(100\%) & 0(0.00\%) \\
    & B2 & 1051(94.7\%) & 1061(95.6\%) & 1055.5(95.1\%) & 3.63(0.33\%) \\
  \end{tabular}
  }
  \label{tab:xvariability}
\end{table}

\begin{table*}[ht]

  \centering
  \caption{\Syntia/ \& \Xyntia/ (\ref{config}): results according to expression
    type and number of inputs (B2, timeout = 60 s)}
  \resizebox{\textwidth}{!}{%
  \begin{tabular}{lcccccccccc}
      & & \multicolumn{3}{c}{Type} & \multicolumn{5}{c}{\# Inputs} &  \\
      \cmidrule(r){3-5}\cmidrule(r){6-10}
      & Property & Bool. & Arith. & MBA & 2 & 3 & 4 & 5 & 6 & All\\
        \midrule
        \multirow{4}{*}{\Syntia/} & Succ. Rate & 53.8\% & 28.6\% & 21.1\% & 77.3\% & 41.0\% & 10.0\% & 2.2\% & 1.1\% & 34.5\% \\
      & Equiv. Range & 53.0\% & 27.8 - 28.1\% & 20.3 - 20.8\% & 74.0 - 75.33\% & 40.3 - 40.5\% & 10.0\% & 2.2\%  & 1.1\% & 33.7 - 34.0\% \\
      & Mean Qual. & 0.53 & 0.61 & 0.71 & 0.57 & 0.60 & 0.67 & 0.12 & 0 & 0.59 \\ \hline
        
      \multirow{4}{*}{\Xyntia/} & Succ. Rate & 98.4\% & 96.5\% & 91.6\% & 98.7\% & 98.8\% & 98.9\% & 82.2\% & 74.4\% & 95.5\% \\
      & Equiv. Range & 97.8\% & 88.9 - 94.9\% & 85.1 - 90.0\% & 93.3 - 97.3\% & 93.2 - 97.7\% & 94.4 - 97.2\% & 80.0 - 81.1\% & 72.2 - 73.3\% & 90.6 - 94.2\% \\
      & Mean Qual. & 0.73 & 1.0 & 1.05 & 0.68 & 0.90 & 1.11 & 0.94 & 1.05 & 0.92 \\
      \midrule
  \end{tabular}
  }
\label{tab:xyntia_inputs_type}
\end{table*}

\myparagraph{\ref{q2}} Unlike \Syntia/, \Xyntia/ is efficient on B2. Moreover, as presented in \cref{tab:xyntia_inputs_type}, it is able to synthesize expressions using up to 5 inputs with a success rate $\geq 80\%$. Even for 6 inputs it still reaches a success rate $>70\%$. Furthermore, it enables to synthesize efficiently different types of expressions. While it seems harder to synthesize MBA expressions, even for \Xyntia/, it still synthesizes $>85\%$ of them. In addition, we observe that \Xyntia/ returns simple and almost always correct results. Still, results given in \cref{tab:xyntia_inputs_type} seems to show that \Syntia/ returns better quality results and less non-equivalent expressions than \Xyntia/. However, these conclusions are biased by the fact that \Syntia/
has a lower success rate than \Xyntia/ and finds only very simple expressions. Thus, we present results on expressions that had been successfully synthesized by both \Syntia/ and \Xyntia/.
\cref{tab:syntia_vs_xyntia_same} demonstrates that under this condition, the quality of 
both tools are comparable. Still, \Xyntia/ reaches such results thanks to our post-process simplifier. Thus, \Syntia/ effectively synthesizes simpler expressions, but the gap can be bridged by adding a simple simplifier to 
\Xyntia/.
On the other hand, we see that \Syntia/ returns between 6 and 9 non-equivalent expressions while \Xyntia/ returns between 1 and 4. Thus \Xyntia/ seems more reliable.

\begin{table}[!ht]
    \centering
    \caption{Results for expressions that both \Syntia/ and \Xyntia/ (\ref{config}) successfully synthesized (B2, timeout = 60 s).}
    \begin{adjustbox}{max width=\columnwidth}
        \begin{tabular}{lccc}
             & \Syntia/ & \Xyntia/ \\
            \cmidrule(r){1-2}\cmidrule(lr){3-3}\cmidrule(lr){4-4}
            \#Succ. & 383 & 383 \\
            \#Equiv. & 374 - 377 & 379 - 382 \\
            Mean Qual. & 0.58 & 0.62 \\

        \end{tabular}
    \end{adjustbox}
    \label{tab:syntia_vs_xyntia_same}
\end{table}

\myparagraph{\ref{q3}}
\Xyntia/ defaults to synthesizing expressions over \EXPR/
while \Syntia/ infers expressions over \FULL/. To evaluate the
sensitivity of \Xyntia/ to search space and show that previous results
was not due to search space inconsistency, we run \Xyntia/ over 
\FULL/, \EXPR/ and \MBA/. \cref{tab:ils_vs_syntia} indicates
that \Xyntia/ reaches high equivalence rates for all operators'
sets -- recall \Syntia/ results stayed low. Still, \Xyntia/ seems more
sensitive to the size of the set of operators than \Syntia/. Its
proven equivalence rate decreases from 90\% (\EXPR/) to 71\%
(\FULL/) while \Syntia/ decreases only from 38.7\% (\EXPR/) to
33.7\% (\FULL/). On the other hand, restricting the search space
to \MBA/ benefits to both \Syntia/ and \Xyntia/.

\begin{table}[ht]
	\centering
        \caption{\Xyntia/ (\ref{config}): results on \FULL/, \EXPR/ and \MBA/ (B2, timeout = 60 s).}
        \begin{adjustbox}{max width=\columnwidth}
        	\begin{tabular}{lcccc}
                	& & \FULL/ & \EXPR/ & \MBA/ \\
                        \cmidrule(r){1-2}\cmidrule(lr){3-3}\cmidrule(lr){4-4}\cmidrule(lr){5-5}
			
			\multirow{4}{*}{\Xyntia/} & Succ. Rate & 85.3\% & 95.5\% & 95.7\% \\
			& Equiv. Range & 71.2 - 76.1\% & 90.6 - 94.2\% & 91.4 - 95.6\% \\
			& Mean Qual. & 1.04 & 0.92 & 0.97 \\
		\end{tabular}
	\end{adjustbox}
	\label{tab:ils_vs_syntia}
\end{table}

\subsubsection{\textbf{Optimal \Xyntia/}}\label{annex:features}

The systematic evaluation of \Xyntia/ depending on some design choices is resumed in \cref{sec:features}. We complete here our analysis and give more details about the measured results. We focus on the following aspects:
\begin{enumerate*}
    \item the choice of the S-metaheuristic; 
    \item the choice of the sampling strategy;
    \item the choice of the distance as objective function;
    \item the effect of our custom simplifier.
\end{enumerate*}

\myparagraph{Choice of the S-metaheuristic}
We compare 5 S-metaheuristics, namely Hill Climbing, Random Walk,
Simulated Annealing, Metropolis Hasting and Iterated Local Search, to find out
the better suited to deobfuscation.  \Cref{tab:success_rate_all_meta} shows
that ILS has a higher equivalence rate than other search heuristics. Moreover, we observe that all S-metaheurstics obtain similar or better results than \Syntia/. The low equivalence rate of Hill Climbing compared to other S\hyph{}metaheuristics can be explained by the fact that it has no way to evade local optimums. Even in this conditions, we observe that its results are not that far from \Syntia/ (which reaches an equivalence rate of $\approx 38\%$ on \EXPR/).
It confirms that estimating non terminal expression's pertinence through simulations as MCTS does is not suitable for deobfuscation (see \cref{sec:syntia_explaination}). It is far more relevant to manipulate terminal expressions only as S-metaheurstics do.

\begin{table}[htbp]
    \centering
    \caption{Synthesis Equivalence Rate for different S-metaheuristics (B2, \ref{config}, timeout = 60 s)}
    \begin{tabular}{@{}lc@{}}
	S-metaheuristic & Equiv. Range \\
	\cmidrule(r){1-1}\cmidrule(lr){2-2}
	Random Walk & 62.3 - 63.4\% \\
	Hill Climbing & 31.9 - 33.1\% \\
	Iterated Local Search & \textbf{90.6 - 94.2\%} \\
	Simulated Annealing & 64.8 - 65.8\% \\
	Metropolis-Hastings & 57.7 - 58.5\%\\ \\
    \end{tabular}
    \label{tab:success_rate_all_meta}
\end{table}

\myparagraph{Effect of the sampling strategy}
\cref{tab:xyntia_nbsamples} presents \Xyntia/'s results for different number of randomly chosen I/O samples. Intuitively, the higher the number of samples is considered,
the more precise is the synthesis specification. Consequently, one may think that increasing the number of samples would negatively impact the success rate and 
positively impact the equivalence rate. The experiments shows that while it improves the equivalence rate (from 74.95\% to 87.39\% for respectively 10 and 100 I/O samples on \EXPR/), 
it does not weaken the success rate. This result can be explained by the fact that more inputs are used by the objetive function to more precisely guide the synthesis. Still, the degree to which the results are impacted depends on the set of operators.
For \MBA/
and \EXPR/, the equivalence range seems to stagnate when adding more
than 50 samples while \FULL/ still improves with 100 samples.

\begin{table}[!ht]
    \centering
    \caption{Results of \Xyntia/  for different number of samples (B2, \ref{config}, timeout = 60 s).}
    \begin{adjustbox}{max width=\columnwidth}
        \begin{tabular}{lcccc}
            \# samples & & \FULL/ & \EXPR/ & \MBA/ \\
            \cmidrule(r){1-2}\cmidrule(lr){3-3}\cmidrule(lr){4-4}\cmidrule(lr){5-5}%\cmidrule(lr){6-6}
            \multirow{4}{*}{10} & Succ. Rate & 85.05\% & 93.69\% & 93.33\% \\
            & Equiv. Range & 52.79 - 59.10\% & 74.95- 79.55\% & 79.64 - 85.14\% \\
            & Mean Qual. & 0.94 & 0.95 & 0.96 \\ \\

            \multirow{4}{*}{20} & Succ. Rate & 86.85\% & 93.96\% & 94.50\% \\
            & Equiv. Range & 59.46 - 65.14\% & 82.61 - 88.65\% & 87.12 - 92.43\% \\
            & Mean Qual. & 1.02 & 0.93 & 0.96 \\ \\

            \multirow{4}{*}{50} & Succ. Rate & 88.65\% & 95.50\% & 96.13\% \\
            & Equiv. Range & 66.49 - 72.34\% & 87.75 - 92.70\% & 89.91 - 95.77\% \\
            & Mean Qual. & 1.04 & 0.92 & 0.96 \\ \\

            \multirow{4}{*}{100} & Succ. Rate & 86.67\% & 95.32\% & 96.58\% \\
	    & Equiv. Range & 69.10 - 75.50\% & 87.39 - 93.51\% & 91.26 - 96.58\% \\
	    & Mean Qual. & 1.05 & 0.94 & 0.95 \\
        \end{tabular}
    \end{adjustbox}
    \label{tab:xyntia_nbsamples}
\end{table}

In order to improve \Xyntia/'s results over the \FULL/ sets of operators we propose to add constant vectors ($\vec{0}, \vec{1}, \vec{-1}, \vec{min_s}, \vec{max_s}$)
to enforce important behaviors such as division by zero and overflows. \cref{tab:xyntia_limits} presents results of \Xyntia/ in two configurations:
\begin{enumerate*}
    \item 100 randomly generated samples and 
    \item 95 randomly generated samples plus 5 constant vectors. 
\end{enumerate*}    
We see that adding such constant vectors slightly improves \Xyntia/'s equivalence rate over the \FULL/ and \EXPR/ sets of operators.  

\begin{table}[!ht]
    \centering
    \caption{\Xyntia/ with and without constant values (B2, \ref{config}, timeout = 60 s).}
    \begin{adjustbox}{max width=\columnwidth}
        \begin{tabular}{lcccc}
            \# samples & & \FULL/ & \EXPR/ & \MBA/ \\
            \cmidrule(r){1-2}\cmidrule(lr){3-3}\cmidrule(lr){4-4}\cmidrule(lr){5-5}%\cmidrule(lr){6-6}
            \multirow{4}{*}{no consts} & Succ. Rate & 86.67\% & 95.32\% & 96.58\% \\
	    & Equiv. Range & 69.10 - 75.50\% & 87.39 - 93.51\% & 91.26 - 96.58\% \\
	    & Mean Qual. & 1.05 & 0.94 & 0.95 \\ \\
            
            \multirow{4}{*}{5 consts} & Succ. Rate & 85.32\% & 95.50\% & 95.68\% \\
			& Equiv. Range & 71.17 - 76.13\% & 90.6 - 94.2\% & 91.35 - 95.59\% \\
			& Mean Qual. & 1.04 & 0.92 & 0.97 \\

        \end{tabular}
    \end{adjustbox}
    \label{tab:xyntia_limits}
\end{table}

\myparagraph{Choice of the distance}
The default design of \Xyntia/ (\cref{sec:xyntia}) leverages the Log-arithmetic distance as objective function. We present in \cref{tab:xyntia_distances} an evaluation of 
\Xyntia/ with the following alternative distances:
\begin{itemize}
    \item $\textrm{Arithmetic}_{\, \vec{o}^{\, *}}(\vec{o}\,) = \underset{i}{\sum} |o_i - o^*_i|$
    \item $\textrm{Hamming}_{\, \vec{o}^{\, *}}(\vec{o}\,) = \underset{i}{\sum}\underset{j=0}{\overset{31}{\sum}} o_{i,j} \oplus o^*_{i,j}$
    \item $\textrm{Xor}_{\, \vec{o}^{\, *}}(\vec{o}\,) = \underset{i}{\sum} o_i \oplus o^*_i$
    \item $\textrm{Log-Arithmetic}_{\, \vec{o}^{\, *}}(\vec{o}\,) = \underset{i}{\sum} \, log_{\,2}(1 + |o_i - o^*_i|)$
\end{itemize}

where $\vec{o}^{\, *}$ is the vector of sampled outputs and $\vec{o}$ is the actual outputs of the synthesized expression.

It appears that the Log-arithmetic distance guides synthesis the best. Over \EXPR/, \Xyntia/ reaches a proven equivalence rate
between 84.50\% with the Hamming distance, and 90.6\% with the Log\hyph{}arithmetic one. While, intuitively, the Xor and Hamming distances should guide the search better for Boolean expressions, \cref{tab:xyntia_distances_types} 
demonstrates that this is not the case: the Log-arithmetic distance is better for Boolean expressions.

\begin{table}[!ht]
    \centering
    \caption{\Xyntia/'s results for different distances (B2, \ref{config}, timeout = 60 s).}
    \begin{adjustbox}{max width=\columnwidth}
        \begin{tabular}{lcccc}
            Dist. & & \FULL/ & \EXPR/ & \MBA/ \\
            \cmidrule(r){1-2}\cmidrule(lr){3-3}\cmidrule(lr){4-4}\cmidrule(lr){5-5}%\cmidrule(lr){6-6}
            \multirow{4}{*}{Arith} & Succ. Rate & 86.58\% & 94.50\% & 95.59\% \\
            & Equiv. Range & 69.55 - 76.22\% & 87.57 - 93.51\% & 89.37 - 95.59\% \\
            & Mean Qual. & 1.14 & 0.98 & 1.01 \\ \\

            \multirow{4}{*}{Hamm.} & Succ. Rate & 83.42\% & 91.53\% & 92.25\% \\
            & Equiv. Range & 67.30 - 73.42\% & 84.50 - 89.73\% & 88.38 - 92.25\% \\
            & Mean Qual. & 1.09 & 0.93 & 0.92 \\ \\

            \multirow{4}{*}{Xor} & Succ. Rate & 82.34\% & 92.34\% & 95.50\% \\
            & Equiv. Range & 66.76 - 73.15\% & 86.07 - 90.09\% & 90.90 - 95.41\% \\
            & Mean Qual. & 1.13 & 0.94 & 0.98 \\ \\

            \multirow{4}{*}{LogArith} & Succ. Rate & 85.32\% & 95.50\% & 95.68\% \\
	    & Equiv. Range & 71.17 - 76.13\% & 90.6 - 94.2\% & 91.35 - 95.59\% \\
	    & Mean Qual. & 1.04 & 0.92 & 0.97 \\
        \end{tabular}
    \end{adjustbox}
    \label{tab:xyntia_distances}
\end{table}

\begin{table}[!ht]
    \centering
    \caption{\Xyntia/'s results for different distances over Boolean and
      arithmetic type of expressions (B2, \ref{config}, timeout = 60 s).}
    \begin{adjustbox}{max width=\columnwidth}
        \begin{tabular}{lccc}
            Dist. & & Boolean & Arith. \\
            \cmidrule(r){1-2}\cmidrule(lr){3-3}\cmidrule(lr){4-4}
            \multirow{4}{*}{Arith} & Succ. Rate & 96.76\% & 96.49\%  \\
            & Equiv. Range & 95.68\% & 87.30 - 95.68\% \\
            & Mean Qual. & 0.75 & 1.05 \\ \\

            \multirow{4}{*}{Hamm.} & Succ. Rate & 97.84\% & 90.81\% \\
            & Equiv. Range & 95.41 - 95.68\% & 81.08 - 89.19\% \\
            & Mean Qual. & 0.76 & 0.98 \\ \\

            \multirow{4}{*}{Xor} & Succ. Rate & 97.84\% & 90.54\% \\
            & Equiv. Range & 96.76 - 97.03\% & 82.16 - 87.03\% \\
            & Mean Qual. & 0.79 & 0.97 \\ \\

            \multirow{4}{*}{LogArith} & Succ. Rate & 98.38\% & 96.48\% \\
	    & Equiv. Range & 97.84\% & 88.11 - 95.14\% \\
	    & Mean Qual. & 0.73 & 0.98 \\
        \end{tabular}
    \end{adjustbox}
    \label{tab:xyntia_distances_types}
\end{table}

\myparagraph{Effect of the simplifier}
\Xyntia/ integrates a simple and efficient simplification engine to post-process the expressions found. The simplification rules, which are partially listed in \cref{tab:simplifier_rules}, are iteratively applied on the expression until a fixpoint is reached. \cref{tab:xyntia_simplifier} presents the quality of synthesized expressions with and without the simplification engine. We observe that the simplifier significantly improves the quality of the expressions and enables us to reach really good quality results (\(\approx\) 1) for \EXPR/ and \MBA/. However, for \FULL/, the quality stays around 1.3. As such, some more engineering might be needed to get better results for \FULL/. Nevertheless, our simplifier efficiently rewrites expressions while adding no significant latency. Indeed, the average time spent with this post-processing step is around 2.6ms.      

\begin{table}[!ht]
  \centering
  \caption{\Xyntia/ quality with and without simplifier (B2, \ref{config}, timeout = 60 s).}
  \begin{adjustbox}{max width=\columnwidth}
  \begin{tabular}{lccc}
        & \FULL/ & \EXPR/ & \MBA/ \\
        \cmidrule(r){2-2}\cmidrule(lr){3-3}\cmidrule(lr){4-4}%\cmidrule(lr){5-5}
        Mean Qual. No Simpl. & 1.77 & 1.33 & 1.38 \\
        Mean Qual. Simpl. & 1.19 & 0.93 & 0.97 \\
        Mean Simpl. Time (s) & 0.0027 & 0.0026 &  0.0026 \\
  \end{tabular}
  \end{adjustbox}
  \label{tab:xyntia_simplifier}
\end{table}

\begin{table}[!ht]
  \centering
  \caption{\Xyntia/'s simplification rules (partial)}
  \begin{adjustbox}{max width=\columnwidth}
  \begin{tabular}{ll@{\hspace{1cm} $\rightarrow$ \hspace{1cm}}l}
        \hline \\
        Constant & $f(const_1, ..., const_N) $ & $result$ \\ \\ \hline \\
        
        \multirow{11}{*}{Arithmetic} & $E_1 + 0$ & $E_1$ \\
        & $E_1 - 0$ & $E_1$ \\
        & $E_1 - (- E_2)$ & $E_1 + E_2$ \\
        & $E_1 - E_1$ & $0$ \\ 
        & $- (- E_1)$ & $E_1$  \\
        & $(- E_1) + E_2$ & $E_2 - E_1$ \\
        & $E_1 \times 0$ & $0$ \\
        & $E_1 \times 1$ & $E_1$ \\
        & $E_1 << 0 $ & $E_1$ \\
        & $E_1 >>_u 0 $ & $E_1$ \\
        & $E_1 >>_s 0 $ & $E_1$ \\ \\ \hline \\
        
        \multirow{10}{*}{Boolean} & $\neg ( \neg E_1 )$ & $E_1$ \\ 
        & $E_1 \land -1$ & $E_1$ \\
        & $E_1 \land 0$ & $0$ \\
        & $E_1 \land E_1$ & $E_1$ \\
        & $E_1 \lor 0$ & $E_1$ \\
        & $E_1 \lor -1$ & $-1$ \\
        & $E_1 \lor E_1$ & $E_1$ \\
        & $E_1 \oplus -1$ & $\neg E_1$ \\
        & $E_1 \oplus 0$ & $E_1$ \\
        & $E_1 \oplus E_1$ & $0$ \\ \\ \hline \\
  \end{tabular}
  \end{adjustbox}
  \label{tab:simplifier_rules}
\end{table}

\subsubsection{\textbf{Limitations}} \label{annex:limitations} 
\begin{figure}[!ht]
\centering
    \resizebox{0.9\columnwidth}{!}{%
    \input{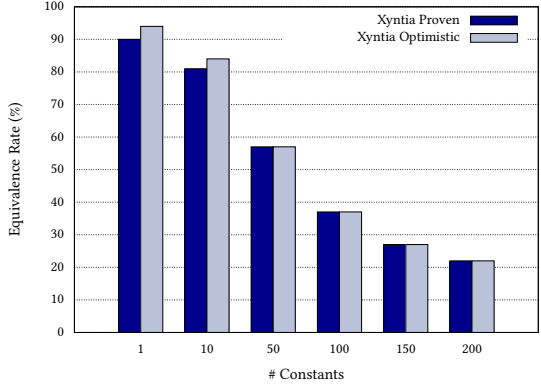}
    }
    \caption{Effect of the number of constant values in \Xyntia/'s grammar on equivalence rate (B2, \ref{config}, timeout=60s)}
\label{fig:nb_consts}
\end{figure}

\cref{sec:limitations} discusses inherent limitations of blackbox methods. While some are extensively
studied in \cref{sec:protections}, we propose to discuss here if AI-based blackbox methods can efficiently synthesize expressions manipulating constant values. Indeed, \Xyntia/ and \Syntia/ only integrate the constant \texttt{1} in their grammar. Thus, if they try to synthesize an expression containing constant values ($\neq 1$) they will need to create them. However, this is unlikely, especially if the constant is far from 1. One solution is to add all $2^{32}$ constant values in the grammar. In order to verify if this approach is conceivable, we add ranges of constant values ($[1; N]$ for $N \in \{1, 10, 50, 100, 200\}$) in \Xyntia/'s grammar. The results for each configuration are presented in \cref{fig:nb_consts}. They show that increasing the number of constant values dramatically impacts \Xyntia/'s performance. We conclude that adding all possible constant values is not beneficial. Another solution is to add well chosen constant values (-1, $min_s$, $max_s$) but we decided not to explore this approach in this paper. Still, in \cref{sec:usecase}, we observe that such restriction is limited as \Xyntia/ is able to synthesize interesting constant values. Note that \Syntia/ cannot do it.

\subsection{Compare to other approaches: More details} \label{annex:compare_other}

We present in \cref{sec:comparative_study} a comparative study of \Xyntia/ against greybox deobfuscators, whitebox simplifiers and state-of-the-art program synthesizers. We give here more details on how \Xyntia/ compares to whitebox simplifiers.

\subsubsection{\textbf{Comparison to whitebox simplifiers}} \label{annex:vswhite}

We compare \Xyntia/ over the EA, VR-EA and EA-ED datasets with 3 whitebox approaches: GCC, Z3 simplifier (v4.8.7) and our custom simplifier.
We use GCC v8.3.0 with optimization level 3 to compile obfuscated expressions. We do not report the mean simplification time of GCC
because our measurements consider the whole compilation process which would not
be fair compared to other methods. We extract expressions' semantics through decompilation for EA and EA-ED datasets (using ghidra \cite{ghidra}) and leverage symbolic execution using Binsec \cite{david2016binsec} for VR-EA dataset where it is not possible to use decompilation. Because, our custom simplification engine does not integrate concatenation simplification, it did not simplify any expressions over VR-EA\footnote{Symbolic execution engine returns expressions with a lot of concatenations}. \cref{tab:whitebox_better} shows that
GCC, Z3 simplify and our custom simplifier
hardly clear expressions compared to \Xyntia/. However, synthesis is on average
slower that syntax based simplifiers.

\begin{table}[htbp]
	\centering
        \caption{Results of whitebox simplifiers on the EA dataset}
        \begin{adjustbox}{max width=\columnwidth}
            \begin{tabular}{lccccc}
                    & & GCC -03 & Simplfier & Z3 & \Xyntia/ \\
                    \cmidrule(r){1-2}\cmidrule(lr){3-3}\cmidrule(lr){4-4}\cmidrule(lr){5-5}\cmidrule(lr){6-6}
                    \multirow{2}{*}{EA} & Enhancement rate & 68 / 500  & 36 / 500 & 22 / 500 & 360 / 500 \\
                    & Mean time (s) & - & 0.005 & 0.0002 & 2.45 \\ \\
                    \multirow{2}{*}{VR-EA} & Enhancement rate & 22 / 500  & 0 / 500 & 31 / 500 & 360 / 500 \\
                    & Mean time (s) & -  & - & 0.0010 & 2.45 \\ \\
                    \multirow{2}{*}{EA-ED} & Enhancement rate & 14 / 500  & 15 / 500 & 17 / 500 & 360 / 500 \\
                    & Mean time (s) & -  & 0.0055 & 0.00042 & 2.45 \\
		\end{tabular}
	\end{adjustbox}
        \label{tab:whitebox_better}
\end{table}

\subsection{Deobfuscation with \Xyntia/} \label{annex:obf}

We show in \cref{sec:deobfuscation_xyntia} that \Xyntia/ bypasses state-of-the-art obfuscation strategies and enables to reverse VM handlers of program obfuscated with Tigress \cite{Tigress} and VMProtect \cite{VMProtect}. We detail now 
\begin{enumerate*}
    \item the obfuscation used in \cref{sec:usual_obf};
    \item scripts to generate Tigress use cases from \cref{sec:usecase}.
\end{enumerate*}

\subsubsection{Effectiveness against usual protections} \label{annex:usual_obf} \cref{sec:usual_obf} shows that \Xyntia/ enables to bypass usual protections. All tested obfuscation, except {\it path-based obfuscation}, were performed through Tigress \cite{Tigress}. \cref{tab:obfcmd} presents the Tigress commands used to generate obfuscated expressions. Conversely, evaluation of path based obfuscation relies on a custom encoding inspired from \cite{ollivier2019kill}. We present it now. 

\textbf{Path-based obfuscation}~\cite{ollivier2019kill,10.1145/2810103.2813663}
takes advantage of the path explosion problem to thwart symbolic execution.
While it is efficient against symbolic based analysis,
what about blackbox ones? The example in \cref{lst:sum_path_obf}, is inspired by the {\bf For} primitive from
\cite{ollivier2019kill}.
It computes the sum of \lstinline{x} and \lstinline{y} adding
loops to increase the number of paths to explore (one path for each value of x and y), effectively killing symbolic execution. However, blackbox deobfuscation 
sees inputs-outputs behaviors only and would successfully synthesize the
expression. To confirm it, we encoded B2 as in
\cref{lst:sum_path_obf}. \cref{tab:xyntia_vs_usual_protect} shows the absence
of impact. 

\begin{lstlisting}[basicstyle=\footnotesize,caption={Sum function with path-oriented obfuscation},label={lst:sum_path_obf}]
int sum(int x, int y){
    int x1, y1;
    for (int i = 0; i < x; i++){
        x1++;
    }
    for (int i = 0; i < y; i++){
        y1++;
    }
    
    return x1 + y1;
}
\end{lstlisting}

\subsubsection{Virtualization based Deobfuscation}
\cref{sec:deobfuscation_xyntia} shows that \Xyntia/ enables to synthesize the VM-handler of software protected with Tigress \cite{Tigress} and VMProtect \cite{VMProtect}. \cref{tab:virtcmd} presents the Tigress commands used to generate the 2 Tigress use cases (note that the ``heavy\_computing'' function contains all mixed boolean-arithmetic expressions and is the one we want to obfuscate). 

\begin{table}[htbp]
	\centering
        \caption{Tigress commands for obfuscation in \cref{sec:usual_obf}}
        \begin{adjustbox}{max width=\columnwidth}
            \begin{tabular}{lp{.9\columnwidth}}
            & Command \\
             \cmidrule(r){2-2}
            MBA & \lstinline[]$tigress --Environment=x86_64:Linux:Gcc:4.6$ \newline \lstinline[]$--Transform=EncodeArithmetic --Functions=fun0$ \newline 
            \lstinline[]$--Transform=EncodeArithmetic --Functions=fun0$ \newline
            \lstinline[]$--out=out.c fun0.c$ \\ \\
            
            Opaque predicate & \lstinline[]$tigress --Environment=x86_64:Linux:Gcc:4.6$ \newline
           \lstinline[]$--Seed=0 --Inputs="+1:int:42,-1:length:1?10"$ \newline
           \lstinline[]$--Transform=InitEntropy --Transform=AddOpaque$ \newline
           \lstinline[]$--Functions=fun0 --AddOpaqueKinds=question$ \newline
           \lstinline[]$--AddOpaqueSplitKinds=inside --AddOpaqueCount=10$ \newline
           \lstinline[]$fun0.c --out=out.c$ \\ \\
            
            Covert channel & \lstinline[]$tigress --Seed=0 --Verbosity=1$ \newline \lstinline[]$--Environment=x86_64:Linux:Gcc:4.6 -pthread$ \newline
           \lstinline[]$--Transform=InitEntropy --Functions=fun0$ \newline
           \lstinline[]$--Transform=InitImplicitFlow --Functions=main$ \newline
           \lstinline[]$--InitImplicitFlowKinds=trivial_thread$ \newline 
           \lstinline[]$--InitImplicitFlowHandlerCount=1$ \newline
           \lstinline[]$--InitImplicitFlowJitCount=1$ \newline 
           \lstinline[]$--InitImplicitFlowJitFunctionBody="(for (if (bb 50) (bb 50)))"$ \newline
           \lstinline[]$--InitImplicitFlowTrace=false --InitImplicitFlowTrain=false$ \newline
           \lstinline[]$--InitImplicitFlowTime=true$ \newline
           \lstinline[]$--InitImplicitFlowTrainingTimesClock=500$ \newline
           \lstinline[]$--InitImplicitFlowTrainingTimesThread=500 fun0.c --out=out.c$ \\
		\end{tabular}
	\end{adjustbox}
        \label{tab:obfcmd}
\end{table}

\begin{table}[htbp]
	\centering
        \caption{Tigress commands for the 2 use cases in \cref{sec:usecase}}
        \begin{adjustbox}{max width=\columnwidth}
            \begin{tabular}{lp{.9\columnwidth}}
            & Command \\
             \cmidrule(r){2-2}
            Tigress (simple) & \lstinline[]$tigress --Environment=x86_64:Linux:Gcc:4.6$ \newline \lstinline[]$--Transform=Virtualize --Functions=heavy_computing$ \newline \lstinline[]$--VirtualizeDispatch=direct --out=out.c main.c$ \\ \\
            Tigress (hard) & \lstinline[]$tigress --Environment=x86_64:Linux:Gcc:4.6$ \newline \lstinline[]$--Transform=EncodeData$ \newline 
            \lstinline[]$--LocalVariables='heavy_computing:v0,v1,v2,v3,v4,v5'$ \newline \lstinline[]$--EncodeDataCodecs=poly1 --Transform=Virtualize$ \newline 
            \lstinline[]$--Functions=heavy_computing --VirtualizeDispatch=direct$ \newline
	        \lstinline[]$--Transform=EncodeArithmetic --Functions=heavy_computing$ \newline
	        \lstinline[]$--Transform=EncodeArithmetic --Functions=heavy_computing$ \newline
	        \lstinline[]$--out=out.c main.c$ \\
                    
		\end{tabular}
	\end{adjustbox}
        \label{tab:virtcmd}
\end{table}

\subsection{Counter AI-based deobfuscation : More details} \label{annex:protections}

Protections against blackbox deobfuscation methods have been discussed extensively in \cref{sec:protections}. We complete in the following
\begin{enumerate*}
    \item the description of the datasets used to 
evaluate the efficiency of the proposed methods;
    \item the results of \Syntia/, CVC4 and STOKE against proposed protections.
\end{enumerate*}

\begin{table}[ht]
	\centering
        \caption{Found expressions on BP1,2,3 (timeout = 1 h)}
        \begin{adjustbox}{max width=\columnwidth}
            \begin{tabular}{lcccc}
                & \Xyntia/ & \Syntia/ & CVC4-\EXPR/ & \stokesynth/ \\
                    \cmidrule(lr){1-1}\cmidrule(lr){2-2}\cmidrule(lr){3-3}\cmidrule(lr){4-4}\cmidrule(lr){5-5}
		        BP1 & 13 & 0 & 0 & 0 \\
		        BP2 & 3 & 0 & 0 & 0 \\
		        BP3 & 1 & 0 & 0 & 0 \\
		\end{tabular}
	\end{adjustbox}
	\label{tab:bps_results_syntia}
\end{table}

\subsubsection{\textbf{Semantically complex handlers}}

\cref{sec:semant_complex} presents an encoding to translate a set of semantically simple handlers to complex ones. The proposed solution enables the creation of handlers as complex as wanted in terms of size and number of arguments.
To evaluate the efficiency of the approach, we created 3 datasets namely BP1 (\cref{tab:bp1}), BP2 (\cref{tab:bp2}) and BP3 (too large to be presented here). Results of \Xyntia/, \Syntia/, CVC4 and \stokesynth/ (\stokeopti/ is not considered as it is not blackbox) are presented in \cref{tab:bps_results_syntia}. It confirms that the protection is highly effective.

\subsubsection{\textbf{Merged handlers}} \label{annex:merged}

\begin{figure}[!ht]
\centering
    \resizebox{\columnwidth}{!}{%
    \input{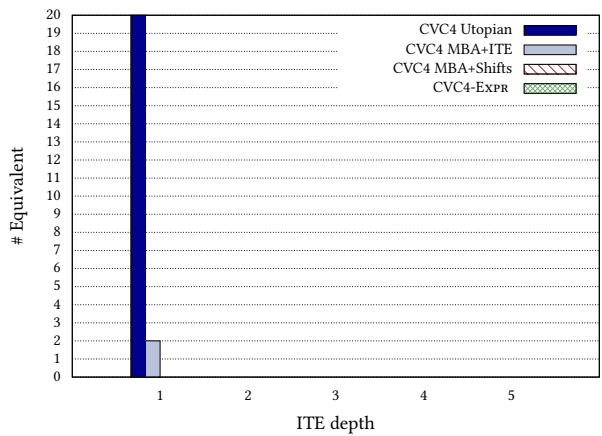}
    }
    \caption{CVC4 against merged handlers (timeout = 60 s)}
\label{fig:cvc4_merged}
\end{figure}

In \cref{sec:merged_handlers}, we measure the impact of conditionals on blackbox methods.
We thus introduce 5 datasets containing 20 expressions each. The first one combines
two basic handlers with one ITE, the second one combines 3 basic handlers though
2 nested conditionals and so on. The first and second datasets are given in
\cref{fig:merged_all}. Each condition compares the third input ($z$) with a
constant. The constant values are always sorted in increasing order starting from
zero -- i.e., the first ITE compares $z$ to 0, the second to 1,
etc.. 

The results of \Xyntia/ against merged handlers are presented in
\cref{sec:merged_handlers}.
We present now the evaluation of \Syntia/, CVC4 and \stokesynth/ (\stokeopti/ is not tested as it is not blackbox) on the same datasets. 
Because we cannot change ITEs to the grammar of \Syntia/ nor \stokesynth/, we evaluate them in their default configuration.
\cref{fig:cvc4_merged} shows that in the utopian configuration CVC4 efficiently synthesizes all expressions with one conditional. However, we see that in any configuration, it is not able to synthesize expressions with nested conditionals. On the other hand, \cref{tab:synt_stoke_merged} shows that neither \Syntia/ nor \stokesynth/ is able to handle merged handler. Results confirm that 
merging handlers is an efficient protection, impeding sampling and synthesis.

\begin{table}[htbp]
	\centering
        \caption{\Syntia/ and \stokesynth/: Number of synthesized expressions against merged handlers (timeout = 60s)}
        \begin{adjustbox}{max width=\columnwidth}
            \begin{tabular}{lccccc}
                    & 1 & 2 & 3 & 4 & 5 \\
                    \cmidrule(r){1-2}\cmidrule(lr){3-3}\cmidrule(lr){4-4}\cmidrule(lr){5-5}\cmidrule(lr){6-6}
                    \Syntia/ & 0  & 0 & 0 & 0 & 0 \\
                    \stokesynth/ & 0  & 0 & 0 & 0 & 0 \\
		\end{tabular}
	\end{adjustbox}
        \label{tab:synt_stoke_merged}
\end{table}

\begin{table*}[htbp]
    \centering
    \caption{Encoding of basic operators (BP1).}
    \begin{adjustbox}{max width=\textwidth}
        \begin{tabular}{ll}
            \cmidrule(r){1-1}\cmidrule(lr){2-2}
            \multirow{3}{*}{$add = h_1 + h_2 + h_3$} & $h_1 = (x + y) + -((a -x^2) - (xy))$ \\
            & $h_2 = (a -x^2) - xy + (- (y - (a \land x)) \times (y \otimes x))$ \\
            & $h_3 = (y - (a \land x)) \times (y \otimes x)$ \\ \\

            \multirow{3}{*}{$sub = h_1 + h_2 + h_3$} & $h_1 = x-y + - (xa - (y \lor a))$ \\
            & $h_2 = (xa - (y \lor a)) + - ((((y \land a) + x) \otimes y) \times x)$ \\
            & $h_3 = (((y \land a) + x) \otimes y) \times x$ \\ \\

            \multirow{3}{*}{$mul = h_1 + h_2 + h_3$} & $h_1 = x \times y + - (xa^2 - xy)$ \\
            & $h_2 = xa^2 - xy + - ((x \otimes y) -(a * (x + y)))$ \\
            & $h_3 = (x \otimes y) -(a * (x + y))$ \\ \\

            \multirow{3}{*}{$and = h_1 \oplus h_2 \oplus h_3$} & $h_1 = x \land y \otimes (x \land a)^2 \lor y$ \\
            & $h_2 = ((x \land a)^2 \lor y) \otimes (ya - ((x \otimes a) + y))$ \\
            & $h_3 = ya - ((x \otimes a) + y)$ \\ \\

            \multirow{3}{*}{$or = h_1 + h_2 + h_3$} & $h_1 = x \lor y + - (ya +x)^2$ \\
            & $h_2 = (ya +x)^2 + - (xa \otimes (y - (x \land a)))$ \\
            & $h_3 = xa \otimes (y - (x \land a))$ \\ \\

        \end{tabular}
    \end{adjustbox}
    \label{tab:bp1}
\end{table*}

\begin{table*}[!ht]
    \centering
    \caption{Encoding of basic operators (BP2).}
    \begin{adjustbox}{max width=\textwidth}
        \begin{tabular}{ll}
            \cmidrule(r){1-1}\cmidrule(lr){2-2}
            \multirow{3}{*}{$add = h_1 + h_2 + h_3$} & $h_1 = (x + y) + ((\neg (xy^2)) \oplus (-a \lor b)) - (x(d \land c))$ \\
            & $h_2 = (-(((\neg (xy^2)) \oplus (-a \lor b)) - (x(d \land c)))) + (((\neg x) \lor (bd)) \land (( a - y - d) \oplus (-c \land (d - x))))$  \\
            & $h_3 = -((\neg x \lor (b \times d)) \land (( a - y - d) \oplus (( -c \land (d - x))))$  \\ \\

            \multirow{3}{*}{$sub = h_1 + h_2 + h_3$} & $h_1 = (x - y) + (((y \oplus c) \times (x \land (-(c \lor b^2)))) - (a + (b \land \neg d)))$  \\
            & $h_2 = (-(((y \oplus c)(x \land (-(c \lor b^2)))) - (a + (b \land \neg d)))) + (((d - (b \lor \neg y)) \land (c + a)) \oplus (x \times - a))$  \\
            & $h_3 = - (((d - (b \lor \neg y)) \land (c + a)) \oplus (x \times - a))$  \\ \\

            \multirow{3}{*}{$mul = h_1 \oplus h_2 \oplus h_3$} & $h_1 = (xy) \oplus (((y + \neg d) \land (x \times a)) \oplus ( -b \lor (c - a)))$  \\
            & $h_2 = (((y + \neg d) \land (x \times a)) \oplus ( -b \lor (c - a))) \oplus (((d \lor -b) \oplus ((x - y)^2)) \land (\neg a + c)) $  \\
            & $h_3 = ((d \lor -b) \oplus ((x - y)^2)) \land (\neg a + c)$  \\ \\

            \multirow{3}{*}{$and = h_1 \oplus h_2 \oplus h_3$} & $h_1 = (x \land y) \oplus (((x + d) \land (y \times (\neg (b \oplus - a)))) - (c \lor b))$  \\
            & $h_2 = (((x + d) \land (y \times (\neg (b \oplus - a)))) - (c \lor b)) \oplus (((d \land b) - (y^2 \lor -a)) \oplus ( x + \neg c))$  \\
            & $h_3 = (((d \land b) - (y^2 \lor -a)) \oplus ( x + \neg c))$  \\ \\

            \multirow{3}{*}{$or = h_1 + h_2 + h_3$} & $h_1 = (x \lor y) + ((((cd) \lor (a - b)) \oplus (-x - \neg y)) \land c)$ \\
            & $h_2 = (-((((cd) \lor (a - b)) \oplus (- x - \neg y)) \land c)) + (((\neg c - (x \lor -b)) \oplus (y + (a \land y))) \times d)$ \\
            & $h_3 = -(((\neg c - (x \lor -b)) \oplus (y + (a \land y))) \times d)$ \\ \\

        \end{tabular}
    \end{adjustbox}
    \label{tab:bp2}
\end{table*}

\begin{figure*}
    \begin{subfigure}[b]{\columnwidth}
    \centering
    \begin{adjustbox}{max width=\textwidth}
        \begin{tabular}{l}
             \multicolumn{1}{c}{Dataset 1} \\
            \cmidrule(r){1-1}
             $ \textrm{if } z = 0 \textrm{ then } x + y \textrm{ else } x - y $ \\    
             $ \textrm{if } z = 0 \textrm{ then } x + y \textrm{ else } x * y $ \\
             $ \textrm{if } z = 0 \textrm{ then } x + y \textrm{ else } x \land y $ \\
             $ \textrm{if } z = 0 \textrm{ then } x + y \textrm{ else } x \lor y $ \\
             $ \textrm{if } z = 0 \textrm{ then } x + y \textrm{ else } x \oplus y $ \\
             $ \textrm{if } z = 0 \textrm{ then } x - y \textrm{ else } x * y $ \\
             $ \textrm{if } z = 0 \textrm{ then } x - y \textrm{ else } x \land y $ \\
             $ \textrm{if } z = 0 \textrm{ then } x - y \textrm{ else } x \lor y $ \\
             $ \textrm{if } z = 0 \textrm{ then } x - y \textrm{ else } x \oplus y $ \\
             $ \textrm{if } z = 0 \textrm{ then } x * y \textrm{ else } x \land y $ \\
             $ \textrm{if } z = 0 \textrm{ then } x * y \textrm{ else } x \lor y $ \\
             $ \textrm{if } z = 0 \textrm{ then } x * y \textrm{ else } x \oplus y $ \\
             $ \textrm{if } z = 0 \textrm{ then } x \land y \textrm{ else } x \lor y $ \\
             $ \textrm{if } z = 0 \textrm{ then } x \land y \textrm{ else } x \oplus y $ \\
             $ \textrm{if } z = 0 \textrm{ then } x \lor y \textrm{ else } x \oplus y $ \\
             $ \textrm{if } z = 0 \textrm{ then } x - y \textrm{ else } x + y $ \\
             $ \textrm{if } z = 0 \textrm{ then } x * y \textrm{ else } x + y $ \\
             $ \textrm{if } z = 0 \textrm{ then } x * y \textrm{ else } x - y $ \\
             $ \textrm{if } z = 0 \textrm{ then } x \land y \textrm{ else } x + y $ \\
             $ \textrm{if } z = 0 \textrm{ then } x \land y \textrm{ else } x - y $ \\

        \end{tabular}
    \end{adjustbox}
   
  \end{subfigure}
  \begin{subfigure}[b]{\columnwidth}
    \centering
    \begin{adjustbox}{max width=\textwidth}
        \begin{tabular}{l}
            \multicolumn{1}{c}{Dataset 2} \\
            \cmidrule(r){1-1}
            $ \textrm{ if } z = 0 \textrm{ then } x + y \textrm{ else } (\textrm{ if } z = 1 \textrm{ then } x - y \textrm{ else } x * y)$ \\
             $ \textrm{ if } z = 0 \textrm{ then } x + y \textrm{ else } (\textrm{ if } z = 1 \textrm{ then } x - y \textrm{ else } x \land y)$ \\
             $ \textrm{ if } z = 0 \textrm{ then } x + y \textrm{ else } (\textrm{ if } z = 1 \textrm{ then } x - y \textrm{ else } x \lor y)$ \\
             $ \textrm{ if } z = 0 \textrm{ then } x + y \textrm{ else } (\textrm{ if } z = 1 \textrm{ then } x - y \textrm{ else } x \oplus y)$ \\
             $ \textrm{ if } z = 0 \textrm{ then } x + y \textrm{ else } (\textrm{ if } z = 1 \textrm{ then } x * y \textrm{ else } x \land y)$ \\
             $ \textrm{ if } z = 0 \textrm{ then } x + y \textrm{ else } (\textrm{ if } z = 1 \textrm{ then } x * y \textrm{ else } x \lor y)$ \\
             $ \textrm{ if } z = 0 \textrm{ then } x + y \textrm{ else } (\textrm{ if } z = 1 \textrm{ then } x * y \textrm{ else } x \oplus y)$ \\
             $ \textrm{ if } z = 0 \textrm{ then } x + y \textrm{ else } (\textrm{ if } z = 1 \textrm{ then } x \land y \textrm{ else } x \lor y)$ \\
             $ \textrm{ if } z = 0 \textrm{ then } x + y \textrm{ else } (\textrm{ if } z = 1 \textrm{ then } x \land y \textrm{ else } x \oplus y)$ \\
             $ \textrm{ if } z = 0 \textrm{ then } x + y \textrm{ else } (\textrm{ if } z = 1 \textrm{ then } x \lor y \textrm{ else } x \oplus y)$ \\
             $ \textrm{ if } z = 0 \textrm{ then } x - y \textrm{ else } (\textrm{ if } z = 1 \textrm{ then } x * y \textrm{ else } x \land y)$ \\
             $ \textrm{ if } z = 0 \textrm{ then } x - y \textrm{ else } (\textrm{ if } z = 1 \textrm{ then } x * y \textrm{ else } x \lor y)$ \\
             $ \textrm{ if } z = 0 \textrm{ then } x - y \textrm{ else } (\textrm{ if } z = 1 \textrm{ then } x * y \textrm{ else } x \oplus y)$ \\
             $ \textrm{ if } z = 0 \textrm{ then } x - y \textrm{ else } (\textrm{ if } z = 1 \textrm{ then } x \land y \textrm{ else } x \lor y)$ \\
             $ \textrm{ if } z = 0 \textrm{ then } x - y \textrm{ else } (\textrm{ if } z = 1 \textrm{ then } x \land y \textrm{ else } x \oplus y)$ \\
             $ \textrm{ if } z = 0 \textrm{ then } x - y \textrm{ else } (\textrm{ if } z = 1 \textrm{ then } x \lor y \textrm{ else } x \oplus y)$ \\
             $ \textrm{ if } z = 0 \textrm{ then } x * y \textrm{ else } (\textrm{ if } z = 1 \textrm{ then } x \land y \textrm{ else } x \lor y)$ \\
             $ \textrm{ if } z = 0 \textrm{ then } x * y \textrm{ else } (\textrm{ if } z = 1 \textrm{ then } x \land y \textrm{ else } x \oplus y)$ \\
             $ \textrm{ if } z = 0 \textrm{ then } x * y \textrm{ else } (\textrm{ if } z = 1 \textrm{ then } x \lor y \textrm{ else } x \oplus y)$ \\
             $ \textrm{ if } z = 0 \textrm{ then } x \land y \textrm{ else } (\textrm{ if } z = 1 \textrm{ then } x \lor y \textrm{ else } x \oplus y)$ \\

        \end{tabular}
    \end{adjustbox}
  \end{subfigure}
  \caption{Merged handlers: datasets 1 and 2}
  \label{fig:merged_all}
\end{figure*}

\end{document}